\documentclass{article}

\usepackage{PRIMEarxiv}
\usepackage{enumitem,kantlipsum}
\newcommand{\overbar}[1]{\mkern 1.5mu\overline{\mkern-1.5mu#1\mkern-1.5mu}\mkern 1.5mu}

\usepackage[utf8]{inputenc} 
\usepackage[T1]{fontenc}    

\usepackage{hyperref}       
\usepackage{xcolor}
\hypersetup{
    colorlinks,
    linkcolor={red!50!black},
    citecolor={blue!50!black},
    urlcolor={blue!80!black}
}

\usepackage{url}            
\usepackage{booktabs}       
\usepackage{amsfonts}       
\usepackage{nicefrac}       
\usepackage{microtype}      
\usepackage{lipsum}
\usepackage{fancyhdr}       
\usepackage{graphicx}       
\usepackage{float}
\graphicspath{{Fig/}}     

\usepackage{natbib}
\usepackage{amsmath}
\usepackage{amssymb}
\usepackage{siunitx}
\DeclareMathOperator*{\argmin}{argmin}

\usepackage{subfig}

\usepackage{siunitx}

\usepackage{appendix}

\usepackage{authblk}

\title{Reciprocity-aware adaptive tile low-rank factorization for large-scale 3D multidimensional deconvolution
}

\author[1]{Fuqiang Chen}
\author[1]{Matteo Ravasi}
\author[2]{David Keyes}
\affil[1]{Earth Science and Engineering Program\protect\\ Physical Science and Engineering Division\protect \\King Abdullah University of Science and Technology (KAUST)}
\affil[2]{Applied Mathematics and Computational Science Program\protect \\
Computer, Electrical and Mathematical Science and Engineering Division\protect \\ King Abdullah University of Science and Technology (KAUST)}
\begin{document}

\maketitle

\begin{abstract}
Low-rank regularization is an effective technique for addressing ill-posed inverse problems when the unknown variable exhibits low-rank characteristics. However, global low-rank assumptions do not always hold for seismic wavefields; in many practical situations, local low-rank features are instead more commonly observed. To leverage this insight, we propose partitioning the unknown variable into tiles, each represented via low-rank factorization.
We apply this framework to regularize multidimensional deconvolution in the frequency domain, considering two key factors. First, the unknown variable, referred to as the Green's function, must maintain symmetry according to the reciprocity principle of wave propagation. To ensure symmetry within the tile-based low-rank framework, diagonal tiles are formulated as the product of a low-rank factor and its transpose if numerically rank-deficient. Otherwise, they are represented by preconditioned dense forms. Symmetry in off-diagonal elements is achieved by parameterizing sub-diagonal tiles as the product of two distinct low-rank factors, with the corresponding super-diagonal tiles set as their transposes.
Second, the rank of the Green's function varies with frequency; in other words, the Green’s function has different ranks at different frequencies. To determine the numerical rank and optimal tile size for each frequency, we first solve the multidimensional deconvolution problem using a benchmark solver. Based on these results, we estimate the optimal tile size and numerical rank for our proposed solver.
We refer to this approach as \textit{reciprocity-aware, factorization-based adaptive tile low-rank regularization}. To further enhance the local low-rank characteristics, we reorder the receivers using a Hilbert space-filling curve, ensuring that receivers closer in physical space are arranged closer in the unknown Green's function matrix. Our method is applied to a large-scale 3D deconvolution problem, showing improved quality compared to the benchmark solver, particularly in noisy data and sparse sampling scenarios.
\end{abstract}


\section*{Introduction}
Seismic data are often complicated by factors such as multiples and back-scattering from overburden structures like salt bodies. To simplify subsequent processing steps, it is crucial to mitigate these effects. Multidimensional deconvolution (MDD) is a technique that can estimate the Green’s function within a target domain through the deconvolution of the down- and up-going components of a wavefield recorded at a specified datum placed just above the domain of interest. This reconstructed observation is free from scattering effects caused by the overburden (i.e., the portion of the medium above the chosen datum including the free surface), and both receivers and virtual sources are relocated onto this datum. MDD has been widely applied in areas such as demultiple \cite[]{Neut2012, RavasiEage2015, boiero2020, Kumaretal2022, Boieroetal2023, Haacke2023}, interferometric redatuming \cite[]{vanderneutetal2011, Wapenaar2011Gji, Vargasetal2021, Shoheietal2011, ravasi2015gji, 9785892}, and imaging \cite[]{Filippoetal2014, Wapenaar2014, Ravasi2016GJI, Joost2017Geo, MatteoandIvan2021Geophysics}. However, MDD is an extremely ill-posed inverse problem due to the band-limited and noisy nature of seismic data, as well as the use of spatially limited acquisition geometries.

Low-rank regularization has shown promise in addressing the ill-posed nature of the MDD problem~\cite[]{Kumaretal2022, Chenetal2023}. This approach assumes that the complex-valued matrices representing the Green's function in the frequency domain are rank-deficient, enabling a reduction in the number of unknowns by eliminating redundant matrix elements. A common method for producing low-rank solutions in inverse problems involves minimizing the nuclear norm of the solution \cite[]{Gu_2014_CVPR}. However, this requires repeated applications of singular value decomposition (SVD) within an iterative scheme, which can be computationally expensive for large-scale MDD problems where the Green's function matrix is high-dimensional. Another way to enforce a low-rank constraint is to directly factorize the unknown matrix into the product of two smaller rectangular matrices \cite[]{Recht2010, Kumaretal2022}, with both sets of entries estimated jointly. This factorization also reduces storage requirements compared to the nuclear norm approach, as it avoids storing a full matrix. Recently, \cite{chen2023physicsaware} extended this idea to produce symmetric, low-rank solutions, which is particularly relevant for MDD as the Green’s function must satisfy physical reciprocity~\citep{Vargasetal2021}. In 2D MDD examples, symmetry not only stabilizes the solving process when the wavefields are contaminated by noise or artifacts but also halves the number of unknowns compared to the asymmetrical low-rank factorization \citep{Kumaretal2022}.

While the Green's function in 3D MDD often exhibits global low-rank characteristics, the degree of rank deficiency can vary significantly across different blocks within each frequency matrix. This variability presents limitations in efficiently representing and storing the Green's function by global low-rank approximation, which is crucial for large-scale computations. In this work, we extend the factorization-based, symmetric low-rank representation to address the specific demands of 3D MDD, where storing dense frequency matrices of the Green's function can require vast amounts of memory—often tens or even hundreds of gigabytes per frequency.

To leverage the inherent low-rank structure, this strategy tiles each frequency matrix and parameterizes each tile as the product of two smaller rectangular matrices. This method allows for the efficient storage of these matrices, as only the low-rank factors need to be stored rather than the dense, full-size matrices. In a similar manner to the 2D MDD case, this low-rank factorization results in significant memory savings, as the memory consumption is proportional to the rank of each tile rather than the dimensions of the full matrix. The lower the rank of the individual tiles, the greater the memory savings and computational efficiency. Thus, a critical question arises: how can we promote and enhance the rank deficiency of these tiles?

In the context of 3D MDD, the column indexes for each row of the matrix representing the unknown Green's function correspond to the response of a virtual source to all receivers across receiver lines in the acquisition setup. The method used to flatten these receivers from different lines into matrix columns is essential for influencing the local ranks of the matrix. Different flattening approaches can result in matrices with significantly varying local ranks.
One common technique arranges the receivers in row-major order, sequentially placing them into matrix columns according to their physical layout. While this approach is straightforward, it has a notable drawback: receivers that are physically close in the acquisition setup may be positioned far apart in the matrix. This separation undermines the ability to exploit spatial correlations, thereby reducing the potential for local low-rank compression.
In contrast, we organize the receivers from different lines into the matrix columns using the Hilbert space-filling curve. This method ensures that geographically close receivers are also close in terms of column indexes within the matrix, preserving spatial relationships and enhancing the potential for local rank deficiency.

Ultimately, the MDD problem, incorporating our proposed tile low-rank regularization, is reformulated as a least squares problem. The objective function is composed of a smooth convex function (typically differentiable) and a convex, potentially non-smooth function. In this formulation, the smooth term often represents the primary loss or objective, aiming to minimize some measure of error or discrepancy, while the non-smooth term typically serves as a regularization component that imposes constraints on the solution, promoting properties such as sparsity or smoothness.
To tackle the challenges associated with large-scale problems and potential non-smoothness in this inverse problem, we employ the Alternating Direction Method of Multipliers (ADMM) \cite[]{8186925}. ADMM is an optimization algorithm that splits the problem into smaller, manageable subproblems, solved iteratively. By leveraging the separable structure of the problem, it combines the benefits of dual ascent and the augmented Lagrangian method to ensure convergence. This alternating update scheme enhances scalability and efficiency, making ADMM particularly effective for large-scale problems with non-smooth objectives or complex constraints.

\section*{Theory}
\begin{figure}
\centering
\includegraphics[width=0.85\textwidth]{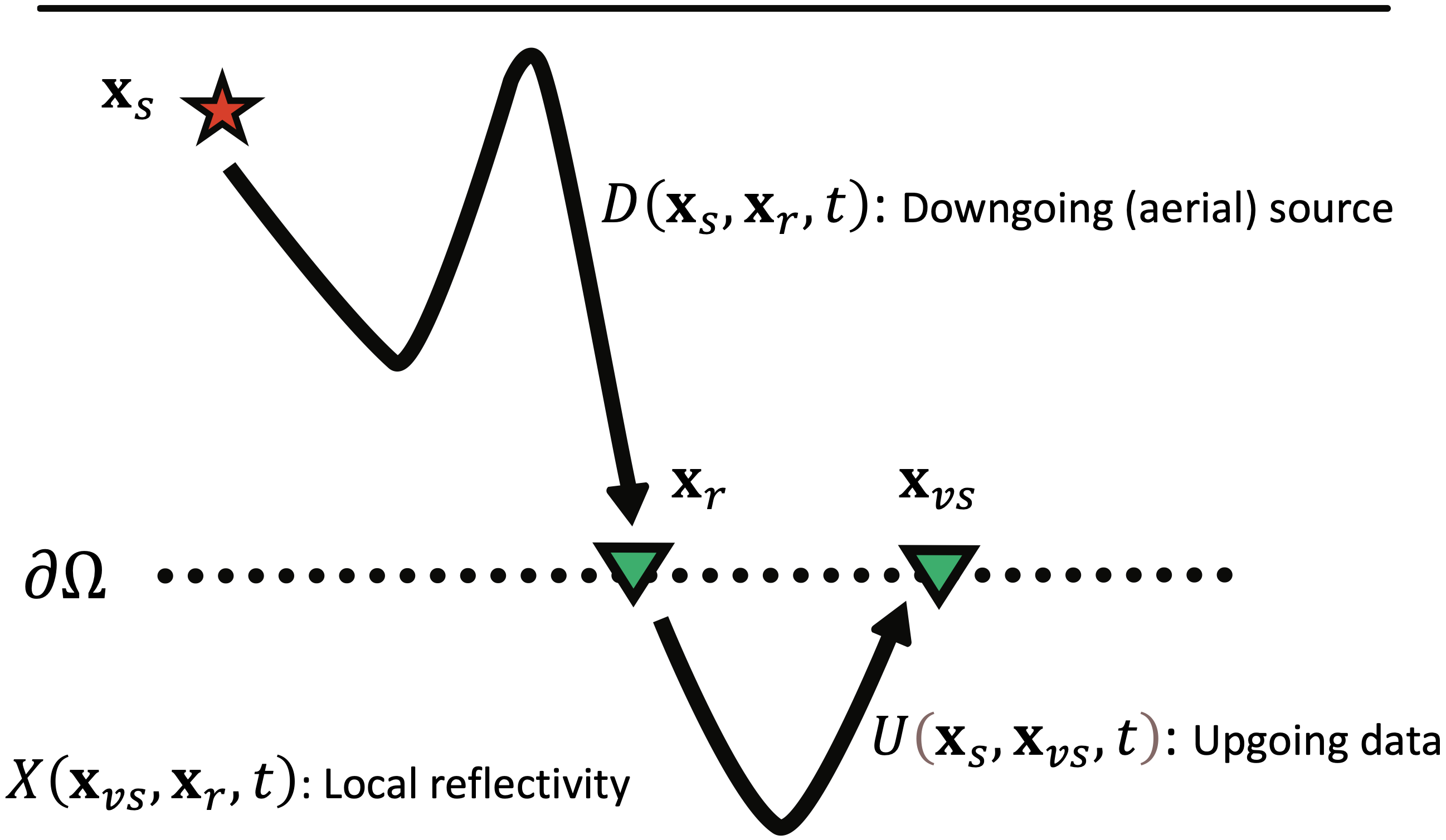}
    \caption{Schematic representation of the acquisition setup considered in MDD problems. A red star represents a source in the overburden above the boundary $\partial \Omega$, while green triangles correspond to receivers (and virtual sources) placed along the boundary of the target area.}
    \label{fig:diagram_mdd}
\end{figure}

Given an acquisition geometry as shown in Figure \ref{fig:diagram_mdd}, associated with a typical MDD application, with sources at locations $\mathbf{x}_s$ and receivers at locations $\mathbf{x}_r$, the down-going $D(\mathbf{x}_s, \mathbf{x}_r,\omega)$ and up-going wavefield $U(\mathbf{x}_s, \mathbf{x}_{vs},\omega) $ are related by the following equation~\citep{amundsen2001, Wapenaar2011Gji}: 
\begin{equation}
U(\mathbf{x}_s, \mathbf{x}_{vs},\omega)=\int_{\partial \Omega } D(\mathbf{x}_s, \mathbf{x}_r,\omega) X(\mathbf{x}_{vs}, \mathbf{x}_r,\omega) d\mathbf{x}_r,
\label{eq:continousMDD_semiFX}
\end{equation}
where $\mathbf{x}_{vs}$ represents the virtual source locations which are positioned along the same boundary as receivers (a common choice in MDD applications). The discrete form of equation \ref{eq:continousMDD_semiFX} can be formulated as:
\begin{equation}
\mathbf{D}\mathbf{X} = \mathbf{U},
\label{eq:axb}
\end{equation}
where $\mathbf{D}\in\mathbb{C}^{n_s\times n_r}$ and $\mathbf{U}\in\mathbb{C}^{n_s\times n_r}$ are matrices representing down- and up-going wavefields, respectively, and $\mathbf{X}\in\mathbb{C}^{n_r\times n_r}$ is the unknown Green's function that we aim to determine. Here, $n_s$ and $n_r$ denote the number of shots and receivers. From a physical standpoint, each row in $\mathbf{D}$ and $\mathbf{U}$ correspond to the response from a single source to all receivers. Similarly, each row in $\mathbf{X}$ represents the response from a virtual source to all receivers. Given the reciprocity principle of wave propagation, $\mathbf{X}$ should be symmetric (i.e., $\mathbf{X}=\mathbf{X}^\mathbf{T}$).
Specifically, the following reciprocity relationship holds for any virtual source-receiver pair: $X(\mathbf{x}_{vs}, \mathbf{x}_r) = X(\mathbf{x}_r, \mathbf{x}_{vs})$. 
We will subsequently refer to this characteristic as the \textit{reciprocity property}. To find the unknown Green's function $\mathbf{X}$, we can solve the minimization problem for each frequency independently as follows:
\begin{equation}
\min_{\mathbf{X}} \frac{1}{2} \Vert\mathbf{D} \mathbf{X} - \mathbf{U}\Vert_\mathcal{F}^2.
\label{eq:standard_solver}
\end{equation}
\subsection{Least-squares preconditioned MDD}
The direct solution of equation \ref{eq:standard_solver} is challenged by issues of non-uniqueness and instability, as in many practical geophysical inverse problems. Addressing these challenges typically requires a combination of strategies, such as incorporating prior physical constraints and introducing a penalty term that enforces smoothness, sparsity or other desirable properties in the solution to stabilize the inverse process. A notable prior assumption about the Green's function $\mathbf{X}$ in MDD is that this matrix is symmetric. However, the solution to equation \ref{eq:standard_solver} does not inherently guarantee this property to hold. To ensure symmetry in the solution, \cite{Vargasetal2021} propose to introduce a symmetrization preconditioner, $\mathbf{P} = 0.5 (\mathbf{I} + \mathbf{\Gamma})$, where $\mathbf{I}$ and $\mathbf{\Gamma}$ are defined such that $\mathbf{I} \mathbf{X} = \mathbf{X}$ and $\mathbf{\Gamma} \mathbf{X} = \mathbf{X}^\mathbf{T}$. Moreover, to further stabilize the inverse problem, we can also apply a commonly used regularization technique to this preconditioned MDD approach, leading to:
\begin{equation}
\min_{\mathbf{Z}} \frac{1}{2}\Vert\mathbf{D} \mathbf{P} \mathbf{Z} - \mathbf{U}\Vert_\mathcal{F}^2 + \lambda \Vert\mathbf{Z}\Vert_\mathcal{F}^2 ,
\label{eq:bench_solver}
\end{equation}
where $\lambda$ is the regularization parameter. After solving equation \ref{eq:bench_solver} to obtain $\mathbf{X}$, the MDD result can be reconstructed by re-applying the preconditioner as $\mathbf{X} = \mathbf{P} \mathbf{Z}$. This approach ensures that the resulting MDD outcomes are both symmetric and exhibit some degree of smoothness. In this paper, we will use this result as a benchmark for comparison against the outcomes of our proposed method.
\begin{figure}[H]
\begin{center}
\includegraphics[width=.5\linewidth]{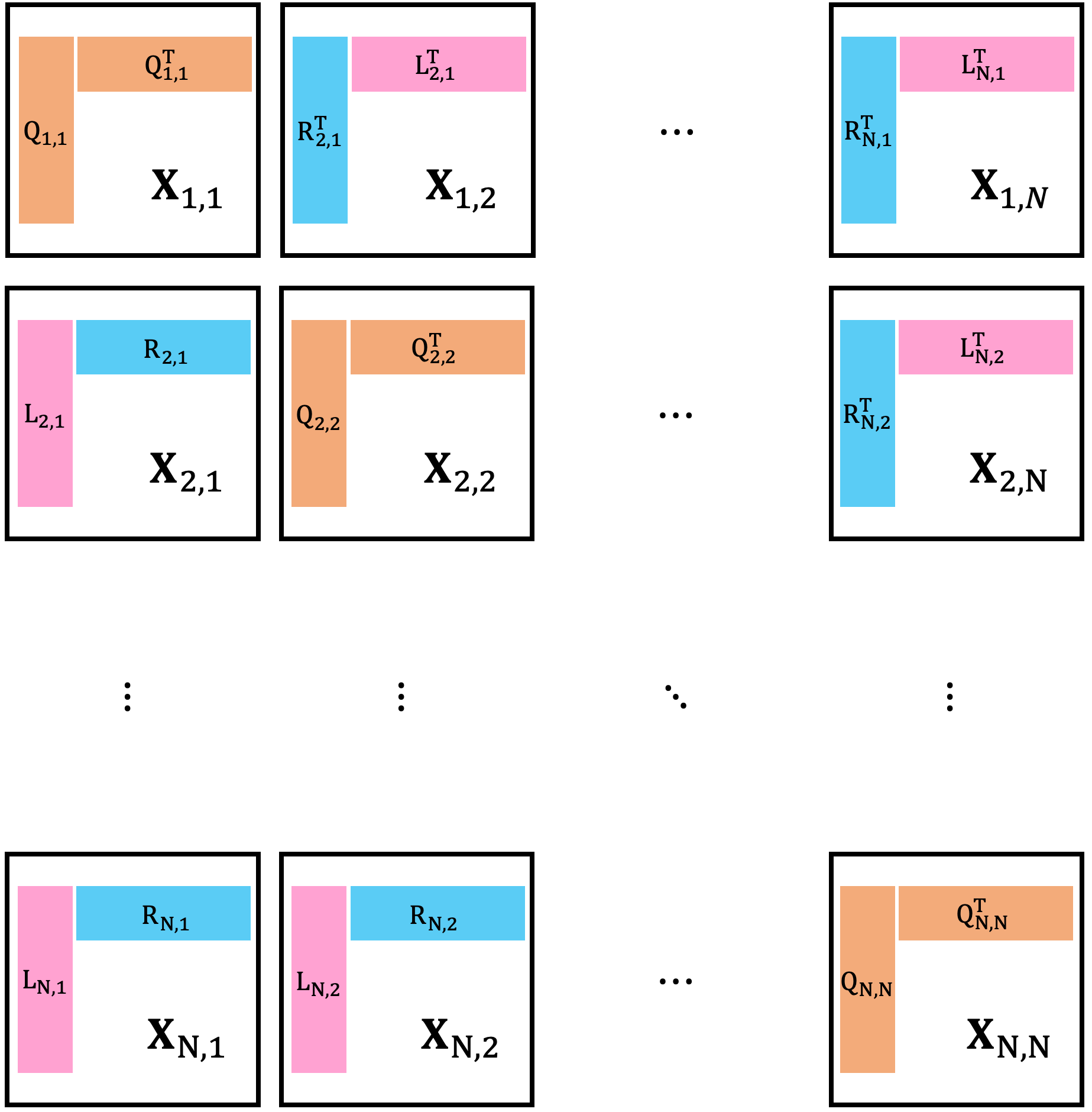}
\caption{A schematic diagram for a $N\times N$ tile low-rank symmetric representation.}
\label{fig:tiledX}
\end{center}
\end{figure}
\subsection{Tile Low-rank MDD}
The benchmark solver in equation \ref{eq:bench_solver} applies Frobenius norm regularization to limit extreme solutions by penalizing large entries on the matrix $\mathbf{X}$. However, one can think of leveraging a more robust prior information for a frequency matrix that represents a seismic wavefield (i.e., $\mathbf{X}\in \mathbb{C}^{n_r\times n_r}$ in our case); namely, as frequently shown in the literature, this matrix tends to show low-rank structure~\citep{stgeo2003,Kreimeretal2013,kumaretal2013eage}. 
In general fitting problems, eigenvectors associated with the largest eigenvalues are often preferred because they capture the dominant data trends (smooth components), while eigenvectors with smaller eigenvalues, which tend to describe fine details, can amplify noise in the inversion process. Low-rank factorization approximations aim to recover the smooth trends by constraining the solution space and shrinking the influence of smaller eigenvalues.
Such a low-rank regularization can be enforced by minimizing the nuclear norm of $\mathbf{X}$. Minimizing the nuclear norm of a matrix encourages sparsity in the singular values of $\mathbf{X}$, and therefore produces a low-rank matrix since most of the singular values are driven down to (or close to) zero. The nuclear norm of a matrix $\mathbf{X}$ is defined as 
$\Vert{\mathbf{X}}\Vert_{\ast}=\sum \sigma({\mathbf{X}})$, where $\sigma({\mathbf{X}})$ denotes the singular values of $\mathbf{X}$. 
By adding a nuclear norm regularization term to the original objective function in equation~\ref{eq:standard_solver}, a new minimization problem can be written as follows 
\begin{equation}
\min_{{\mathbf{X}}} (1/2)\Vert{\mathbf{D}}{\mathbf{X}}-{\mathbf{U}}\Vert_\mathcal{F}^2+\lambda\Vert{\mathbf{X}}\Vert_\ast,
\label{eq:mdd_rd_lsq_reg}
\end{equation}
where $\lambda$ balances the minimization of the data fitting error with enforcing the nuclear norm of $\mathbf{X}$.
Solving equation \ref{eq:mdd_rd_lsq_reg} using iterative solvers requires repeated computations of the SVD of $\mathbf{X}$ \citep{Recht2010, Cai2010,toh2010accelerated}. This process can become computationally expensive, particularly when the matrix $\mathbf{X}$ is large.

An alternative approach to promoting a low-rank solution in the MDD problem is to leverage a low-rank factorization of the solution, which eliminates the need for SVD. Specifically, the matrix $\mathbf{X}$ is approximated by the product of two low-rank matrices, $\mathbf{L}$ and $\mathbf{R}$, as $\mathbf{X} \approx \mathbf{LR}$ \cite[]{kumaretal2013eage}, where $\mathbf{L}\in \mathbb{C}^{n_r\times n_k}$, $\mathbf{R}\in \mathbb{C}^{n_k\times n_r}$, and $n_k$ is the chosen rank for the original matrix $\mathbf{X}$, typically much smaller than $n_r$. This approach achieves the desired low-rank regularization by directly working with the factorized form of $\mathbf{X}$.

\cite{chen2023physicsaware} proposed another factorization for MDD, given by $\mathbf{X} = \mathbf{Q}\mathbf{Q}^\mathbf{T}$, where $\mathbf{T}$ denotes the transpose. This formulation not only ensures low-rank property but also enforces symmetry, which further stabilizes the solution. Compared to the factorization $\mathbf{X}=\mathbf{LR}$, it offers additional memory efficiency since only the factor $\mathbf{Q}$ needs to be stored, rather than both $\mathbf{L}$  and $\mathbf{R}$.
From a performance perspective, the reduced memory footprint in low-rank representations means that computations can be performed at higher levels of the memory hierarchy. This reduces data transfer costs, which often dominate computational expenses in modern hardware. Although this symmetrical low-rank factorization has demonstrated robustness in 2D MDD, particularly when the down-going wavefield contains strong coherent noise \cite[]{chen2023physicsaware}, such a global low-rank factorization approach is not well-suited for 3D MDD, as shown in a subsequent numerical example. 

In this paper, we propose a local low-rank factorization method, as shown in Figure~\ref{fig:tiledX}. The matrix \(\mathbf{X}\) is divided into \(N \times N\) tiles, where each tile is a block of matrix \(\mathbf{X}\). We categorize these blocks into three types:
\begin{enumerate}[leftmargin=*]
    \item \textbf{Diagonal Tiles:}
    
These blocks are represented as \(\mathbf{X}_{i,i}\) for \(i \in (1, \cdots, N)\). If a diagonal block exhibits low-rank structure, then this tile is approximated using low-rank factor \(\mathbf{Q}_{i,i}\) as:
\begin{equation}
\mathbf{X}_{i,i} = \mathbf{Q}_{i,i} \mathbf{Q}_{i,i}^\mathbf{T}.
\end{equation}
Otherwise, we have 
\begin{equation}
    \mathbf{X}_{i,i}=\mathbf{P}\mathbf{Z}_{i,i}.
\end{equation}
Here, each diagonal tile has dimensions of \(d_i \times d_i\), with $\sum\limits_{i=1}^N d_i=n_r$, the low-rank factor $\mathbf{Q}_{i,i}$ has the dimension of $d_i\times k_{i,i}$ with $k_{i,i}$ denotes the rank of the tile $\mathbf{X}_{i,i}$, and the full-rank tile $\mathbf{Z}_{i,i}$ denotes the un-factorized form of the block. It is worth noting that, in this paper, we partition $\mathbf{X}$ for a given frequency into tiles of the same size, even though this is not a necessary constraint.

\item \textbf{Sub-Diagonal Tiles:}

Denoted as \(\mathbf{X}_{i,j}\) for \(i > j\) with \(i \in (2,\dots, N)\) and \(j \in (1,\dots, i-1)\), these tiles are approximated using low-rank factors \(\mathbf{L}_{i,j}\) and \(\mathbf{R}_{i,j}\) as:
\begin{equation}
    \mathbf{X}_{i,j} = \mathbf{L}_{i,j} \mathbf{R}_{i,j}
\end{equation}
The tile $\mathbf{X}_{i,j}$ has the dimension of \(d_i \times d_j\).
The low-rank factors $\mathbf{L}_{i,j}$ and $\mathbf{R}_{i,j}$ have the dimension of $d_i\times k_{i,j}$ and $k_{i,j}\times d_j$, respectively. Here $k_{i,j}$ denotes the rank of the tile $\mathbf{X}_{i,j}$.
\item \textbf{Super-Diagonal Tiles:}

Represented as \(\mathbf{X}_{j,i}\) for \(i > j\) with \(i \in (2,\dots, N)\) and \(j \in (1,\dots, i-1)\), we define them as:
        \begin{equation}
        \mathbf{X}_{j,i} = \mathbf{X}_{i,j}^\mathbf{T}
        \end{equation}
The tile $\mathbf{X}_{j,i}$ has a dimension of \(d_j \times d_i\).
\end{enumerate}
With the representation above for all tiles, the dense matrix $\mathbf{X}$ is symmetric to satisfy the reciprocity principle. Now when we solve the MDD problem, the unknown is not the dense matrix, instead the unknowns will be low-rank factors: $\mathbf{Q}_{i,i}$, $\mathbf{L}_{i,j}$, and $\mathbf{R}_{i,j}$ if all tiles are assumed to be rank-deficient, or low-rank factors combined with a dense representation for the diagonal tiles: $\mathbf{Q}_{i,i}$/$\mathbf{Z}_{i,i}$, $\mathbf{L}_{i,j}$, and $\mathbf{R}_{i,j}$. To deal with the solver discussed below, we introduce new matrices $\mathbf{\overbar Z}$, $\mathbf{\overbar L}$, and $\mathbf{\overbar R}$ defined by stacking the individual tiles:
\begin{itemize}
    \item \(\overbar{\mathbf{Z}} = (\mathbf{Q}_{1,1}/\mathbf{Z}_{1,1}, \mathbf{Q}_{2,2}/\mathbf{Z}_{2,2}, \cdots,\mathbf{Q}_{N,N}/\mathbf{Z}_{N,N})\) for diagonal tiles.
    \item \(\overbar{\mathbf{L}} = (\mathbf{L}_{2,1}, \mathbf{L}_{3,1}, \cdots,\mathbf{L}_{N,N-1})\) for off-diagonal tiles.
    \item \(\mathbf{\overbar R} = (\mathbf{R}_{2,1}, \mathbf{R}_{3,1}, \cdots,\mathbf{R}_{N,N-1})\) for off-diagonal tiles.
\end{itemize}

As we can see, to represent the tiles above, we need information about both the tile size and the rank of each tile. However, a challenge arises because the matrix $\mathbf{X}$ is initially unknown, making the individual blocks $\mathbf{X}_{i,j}$ inaccessible. As a result, the rank of these blocks cannot be determined in advance. To address this challenge, we propose that an estimate of $\mathbf{X}$ be obtained using a benchmark solver, from which the tile size and numerical rank can be inferred. More details are presented in the section on hyperparameter selection and initialization.

\subsection{Receivers re-ordering}
A fundamental assumption underlying the tile low-rank representation described in the previous section is that most of the tiles possess a low-rank structure, characterized by rapidly decreasing singular values.
Though the singular values of the dense matrix are invariant with respect to permutation of its rows or columns, the singular value distribution of the individual tiles of a matrix can be substantially affected by such a permutation. More specifically, when this matrix represents a frequency-domain seismic wavefield, rows and columns are related to the physical sources and receivers (or virtual sources, depending on which matrix in equation~\ref{eq:axb} is considered).

As previously shown in \cite{ravasihongetalimage2022} for the operator matrix (i.e., $\mathbf{D}$), ordering sources along sailing lines and receivers along cables (what we refer from here onwards as the \textit{natural} ordering of seismic data) may not yield optimally compressible blocks. On the other hand, re-organizing the rows and columns of the matrix based on the Hilbert space-filling curve can provide an significant improvement in the compressibility of the different tiles. In other words, by applying Hilbert ordering we ensure that pairs of sources (and receivers) that are close to each other in geographical coordinates are also placed near each in terms of rows (and columns) of the corresponding data matrix (and so, sequentially laid out computer memory); likewise, for pairs of sources (or receivers) that are further apart in geographical coordinates, the corresponding data is also placed in non-nearby rows (and columns) of the corresponding matrix. By so doing, tiles that are closer to the main diagonal of the matrix contain that from sources and receivers that are close to each other, while all other tiles contain data from groups of sources and receivers that are far apart: this is likely to induce faster singular value decay for most of the tiles. A similar argument can also be applied to the unknown matrix $\mathbf{X}$, where receivers and virtual sources are now associated with the rows and columns of the matrix \cite[]{fuqiangetal2024image}.

In this work, we employ this re-ordering approach, mathematically expressed as:
\begin{equation}
\mathbf{S}^\mathbf{T}\mathbf{X}\mathbf{S} = \mathbf{X}_h,
\label{eq:hilbert}
\end{equation}
where the unitary matrix $\mathbf{S}$ represents the re-ordering matrix, and $\mathbf{XS}$ and $\mathbf{S}^\mathbf{T}\mathbf{X}$ denote two matrices with re-arranged columns and rows based on $\mathbf{S}$. First of all, we note that if $\mathbf{X}$ is symmetric, applying the same re-ordering to rows and columns is guaranteed to produce a new matrix $\mathbf{X}_h$ that is also symmetric. 

Next, using the fact that $\mathbf{S}^T = \mathbf{S}^{-1}$, we can write an equivalent expression from equation~\ref{eq:hilbert}
\begin{equation}
\mathbf{X} = \mathbf{S} \mathbf{X}_h \mathbf{S}^\mathbf{T},
\label{eq:hilbert1}
\end{equation}
and substitute this into equation \ref{eq:axb}:
\begin{equation}
\mathbf{D}\mathbf{S}\mathbf{X}_h\mathbf{S}^\mathbf{T} = \mathbf{U}.
\label{eq:axb_2}
\end{equation}
Multiplying now $\mathbf{S}$ on both sides of equation \ref{eq:axb_2}, we obtain 
\begin{equation}
\mathbf{D}\mathbf{S}\mathbf{X}_h = \mathbf{U}\mathbf{S}.
\label{axb}
\end{equation}
If we finally let $\mathbf{\widehat{D}}=\mathbf{DS}$ and $\mathbf{\widehat{U}}=\mathbf{US}$, we obtain an equivalent MDD expression for $\mathbf{X}_h$
\begin{equation}
\mathbf{\widehat{D}}\mathbf{X}_h = \mathbf{\widehat{U}}.
\label{eq:axb_sd}
\end{equation}
In summary, applying the Hilbert curve-based re-ordering to the down- an up-going wavefield matrices $\mathbf{D}$ and $\mathbf{U}$, respectively, involves creating new matrices $\mathbf{DS}$ and $\mathbf{US}$, by re-ordering the columns of $\mathbf{D}$ and $\mathbf{U}$. These reordered matrices are denoted as $\widehat{\mathbf{D}}$ and $\widehat{\mathbf{U}}$. After solving equation \ref{eq:axb_sd} for the unknown $\mathbf{X}_h$, we can finally transform this back to $\mathbf{X}$ using equation \ref{eq:hilbert1}, allowing us to retrieve the Green's function as the MDD result ordered naturally. For simplicity, we will continue to use $\mathbf{D}$ and $\mathbf{U}$ to represent $\widehat{\mathbf{D}}$ and $\widehat{\mathbf{U}}$ in subsequent sections, as no ambiguity arises.

\subsection{ADMM solver}
With the proposed tile parametrization that naturally honors wavefield reciprocity, we formally express the relationship from tiles to dense matrix $\mathbf{X}$ as:
\begin{equation}
    \mathbf{X} = \mathcal{C}(\mathbf{\overbar Z}, \mathbf{\overbar L}, \mathbf{\overbar R}).
\end{equation} 
We then formulate the MDD inverse problem as follows:
\begin{equation}
    \argmin\limits_{\mathbf{X}}\,f(\mathbf{X})+g(\mathbf{X}),
    \label{eq:X2LQR_lst}
\end{equation}
where
\begin{subequations}
\begin{align}
f(\mathbf{X}) &= 0.5\|\mathbf{D}\mathcal{C}(\mathbf{\overbar Z}, \mathbf{\overbar L}, \mathbf{\overbar R})-\mathbf{U}\|_\mathcal{F}^2,\\
g(\mathbf{X})&=g_z(\mathbf{\overbar Z})+g_l(\mathbf{\overbar L})+g_r(\mathbf{\overbar R}).
\end{align}
\end{subequations}
Here $f(\mathbf{X})$ is the smooth coupling function represents the data misfit,
while $g_z(\mathbf{\overbar Z})$, $g_l(\mathbf{\overbar L})$, and $g_r(\mathbf{\overbar R}$) act as extra regularization terms on different low-rank factors \cite[]{Recht2010}. In case these extra regularization terms are differentiable, nonlinear conjugate gradient methods or quasi-Newton methods are suitable for solving the problem. Otherwise, proximal operator-based methods are commonly efficient and well-suited for handling such regularization terms. In our paper, we choose the ADMM approach. Similar to accelerated proximal gradient methods, ADMM naturally handles non-smooth regularization terms through proximal operators without requiring gradient approximations or smoothing. Furthermore, ADMM decomposes the optimization problem into smaller subproblems that can be solved independently, making it scalable for large datasets and distributed computation. Compared to other solvers based on proximal gradient methods, we observed that ADMM achieves faster convergence without the need for additional tuning, as it leverages the quasi-Newton method to efficiently solve the optimization problem related to the smooth data misfit. To solve the optimization problem in equation \ref{eq:X2LQR_lst} with the ADMM, we reformulate the problem as:
\begin{equation}
    \min_{\mathbf{X}, \mathbf{Z}} f(\mathbf{X})+g(\mathbf{Z}), \quad \text{subject to } \mathbf{X} = \mathbf{Z}.
\end{equation}
The ADMM algorithm alternates between updating the primal variables $\mathbf{X}$ and $\mathbf{Z}$ and the dual variable $\Lambda$, using the following steps:
\begin{enumerate}[leftmargin=*]
    \item \textbf{$\mathbf{X}$-update:}
    \begin{equation}
            \mathbf{X}^{k+1} = \arg\min_\mathbf{X} \left( f(\mathbf{X}) + \frac{\rho}{2} \|\mathbf{X} - \mathbf{Z}^k + \Lambda^k\|_\mathcal{F}^2 \right),
    \end{equation}
    where $\rho > 0$ is the penalty parameter. The $\mathbf{X}$-update typically involves minimizing $f(\mathbf{X})$ with a quadratic penalty term. We use limited-memory BFGS \citep{Liu1989} for this update.
    \item \textbf{$\mathbf{Z}$-update:}
    \begin{equation}
    \mathbf{Z}^{k+1} = \arg\min_\mathbf{Z} \left( g(\mathbf{Z}) + \frac{\rho}{2} \|\mathbf{X}^{k+1} - \mathbf{Z} + \Lambda^k\|_\mathcal{F}^2 \right).      
    \label{eq:zupdate}
    \end{equation}
    The $\mathbf{Z}$-update involves computing the proximal operator of $g(\mathbf{Z})$. Let us first define the proximal operator of a generic function $g(\mathbf{Z})$ as in \cite{FISTA2009}:
\begin{equation}
    \operatorname{prox}_g(\mathbf{V})=\argmin\limits_\mathbf{Z}{\big(g(\mathbf{Z})+(1/2)\|\mathbf{Z}-\mathbf{V}\|_\mathcal{F}^2\big)}.
    \label{eq:proxdef}
\end{equation}
If we choose $g_z(\mathbf{\overbar Z})=\lambda\|\mathbf{\overbar Z}\|_\mathcal{F}^2$, $g_l(\mathbf{\overbar L})=\lambda\|\mathbf{\overbar L}\|_\mathcal{F}^2$, and $g_r(\mathbf{\overbar R})=\lambda\|\mathbf{\overbar R}\|_\mathcal{F}^2$ as proposed in \cite{Recht2010}, where the hyperparameter $\lambda$ controls the strength of the regularization terms, the corresponding proximal operator has a closed-form solution:
\begin{equation}
    \operatorname{prox}_{\lambda\|\cdot\|_\mathcal{F}^2}(\mathbf{V})=\frac{\mathbf{V}}{1+2\lambda}.
    \label{eq:proxofFro}
\end{equation}
Then, the $\mathbf{Z}$-update in equation \ref{eq:zupdate} becomes
\begin{equation}
    \mathbf{Z}^{k+1}=\frac{\mathbf{X}^{k+1}+\Lambda^k}{1+2\lambda/\rho}.
    \label{eq:proxofFro2}
\end{equation}
    \item \textbf{Dual variable update:}
    \begin{equation}
    \Lambda^{k+1} = \Lambda^k + \rho (\mathbf{X}^{k+1} - \mathbf{Z}^{k+1}).
    \end{equation}
    This step adjusts the dual variable $\mathbf{\Lambda}$ by incorporating the scaled residual of the constraint, thereby aligning the primal variables $\mathbf{X}$ and $\mathbf{Z}$ over iterations. It ensures convergence to a solution that satisfies the constraints or the original optimization problem.
\end{enumerate}
In summary, this decomposition in ADMM updates allows the two components of the objective function to be addressed independently, making ADMM efficient for problems where $f(\mathbf{X})$ is smooth and $g(\mathbf{X})$ is non-smooth.
Under standard convexity assumptions on $f(x)$ and $g(x)$, the ADMM algorithm is guaranteed to converge to a global solution. Its modular structure also makes it attractive for distributed or large-scale optimization problems. We use the ADMM program provide in \citet*{Ravasi2024PyProximalS} as the solver.

The last ingredient of this algorithm involves computing the derivatives of the data misfit term with respect to all of the different low-rank factors and dense tiles used to represent $\mathbf{X}$. Let $\mathbf{W}=\mathbf{D}^\mathrm{H}(\mathbf{DX}-\mathbf{U})$ denote the gradient of $f$ with respect to $\mathbf{X}$, and $\mathbf{W}'=\mathbf{W}+\mathbf{W}^\mathbf{T}$. To obtain the gradient of $f$ with respect to the off-diagonal tiles, we partition $\mathbf{W}'$ in the same form as $\mathbf{X}$, yielding:
\begin{subequations}
\label{eq:DXLR}
\renewcommand{\theequation}{\theparentequation.\arabic{equation}}
\begin{alignat}{2}
 \nabla_{\mathbf{L}_{i,j}} f =&\mathbf{W}'_{i,j}\mathbf{R}_{i,j}^\mathrm{H},\label{eq:dXl} \\
   \nabla_{\mathbf{R}_{i,j}} f =&\mathbf{L}_{i,j}^\mathrm{H}\mathbf{W}'_{i,j},\label{eq:DXR}
\end{alignat}
\label{eq:grad}
\end{subequations}
\noindent
where $\mathbf{W}'_{i,j}$ corresponds to the partitioned blocks of $\mathbf{W}'$.
For the diagonal tiles represented with the full rank, we obtain: 
\begin{equation}
      \nabla_{\mathbf{Z}_{i,i}} f = 0.5\mathbf{W}'_{i,i}.
\end{equation}
If these tiles are represented with the low-rank representation, we have 
\begin{equation}
      \nabla_{\mathbf{Q}_{i,i}} f = \mathbf{W}'_{i,i}\overbar{\mathbf{Q}}_{i,i},
\end{equation}
where $\overbar{\mathbf{Q}}_{i,i}$ is the conjugate of ${\mathbf{Q}}_{i,i}$.

\subsection{Hyper-parameter selection and initialization}
Our algorithm involves three key hyperparameters: $\lambda$, which controls the regularization for low-rank factors, the tile dimensions, and the assumed rank for each tile. To determine the value of $\lambda$, we employ an iterative procedure where $\lambda_{i+1} = 0.1 \times \lambda_i$. Starting with $\lambda_0 = 10$, the solver is run for three stages, resulting in $\lambda_2 = 0.1$ as the final value.

Additionally, the down-going and up-going wavefields are normalized through the following steps. First, we compute the maximum spectral value of the down-going wavefield, defined as
$\alpha = \max\left( |D(\omega_1)|, |D(\omega_2)|, \dots, |D(\omega_{n_\omega})| \right).
$
Both wavefields are then scaled by $\alpha$, such that $D(\omega) \gets D(\omega)/\alpha$ and $U(\omega) \gets U(\omega)/\alpha$.

To determine the optimal tile size, we first compute the MDD solution $\overbar{\mathbf{X}}$ using the benchmark solver for each frequency. A list of candidate tile sizes is considered, and the solution $\overbar{\mathbf{X}}$ is partitioned accordingly. For each tile, we perform SVD and truncate the singular values based on a predefined threshold to estimate the numerical rank of each tile. The optimal tile size and corresponding ranks are chosen to minimize the number of unknown parameters required to represent the dense matrix $\overbar{\mathbf{X}}$. Further details are provided in the examples section.

Once the optimal tile size and ranks are determined, a warm start is constructed to initialize the proposed method. Specifically, the solution $\overbar{\mathbf{X}}$ from the benchmark solver is tiled using the optimal tile size and ranks, with each tile denoted by $\overbar{\mathbf{X}}_{i,j}$. For off-diagonal tiles ($i \neq j$), SVD is performed:
\begin{equation}
\mathbf{U}_{i,j}, \mathbf{\Sigma}_{i,j}, \mathbf{V}_{i,j} = \mathrm{SVD}\left(\mathbf{X}_{i,j}\right).    
\end{equation}
The results are truncated to retain only the top $k$ singular values and their associated singular vectors:
\begin{subequations}
\begin{align}
\mathbf{\Sigma}_{i,j} &\gets \mathbf{\Sigma}_{i,j}[:k_{i,j}, :k_{i,j}], \\
\mathbf{U}_{i,j} &\gets \mathbf{U}_{i,j}[:, :k_{i,j}], \\
\mathbf{V}_{i,j} &\gets \mathbf{V}_{i,j}[:, :k_{i,j}],
\end{align}
\end{subequations}
where $k_{i,j}$ is the numerical rank corresponding to tile $(i,j)$ with the optimal tile size. The warm initialization is given by:
\begin{subequations}
\begin{align}
\mathbf{L}_{i,j} &= \mathbf{U}_{i,j}\mathbf{\Sigma}_{i,j}^{\frac{1}{2}}, \\
\mathbf{R}_{i,j} &= \mathbf{\Sigma}_{i,j}^{\frac{1}{2}}\mathbf{V}_{i,j}.
\end{align}
\end{subequations}
For the diagonal blocks $\mathbf{X}_{i,i}$, the warm low-rank initialization $\mathbf{Q}_{i,i}$ is obtained by solving:
\begin{equation}
    \mathbf{Q}_{i,i}\mathbf{Q}_{i,i}^T = \mathbf{X}_{i,i}.
    \label{eq:qqinit}
\end{equation}
The cost for solving equation \ref{eq:qqinit} is computationally negligible compared to solving the full MDD problem. If the diagonal tiles are represented in dense form, the initialization is directly set as:
\begin{equation}
    \mathbf{Z}_{i,i} = \mathbf{X}_{i,i}.
\end{equation}

\section*{Numerical Examples}
\subsection{3D Survey Setup and Data Overview}
\begin{figure}
     \centering
     \subfloat[][]{\includegraphics[width=.4\linewidth]{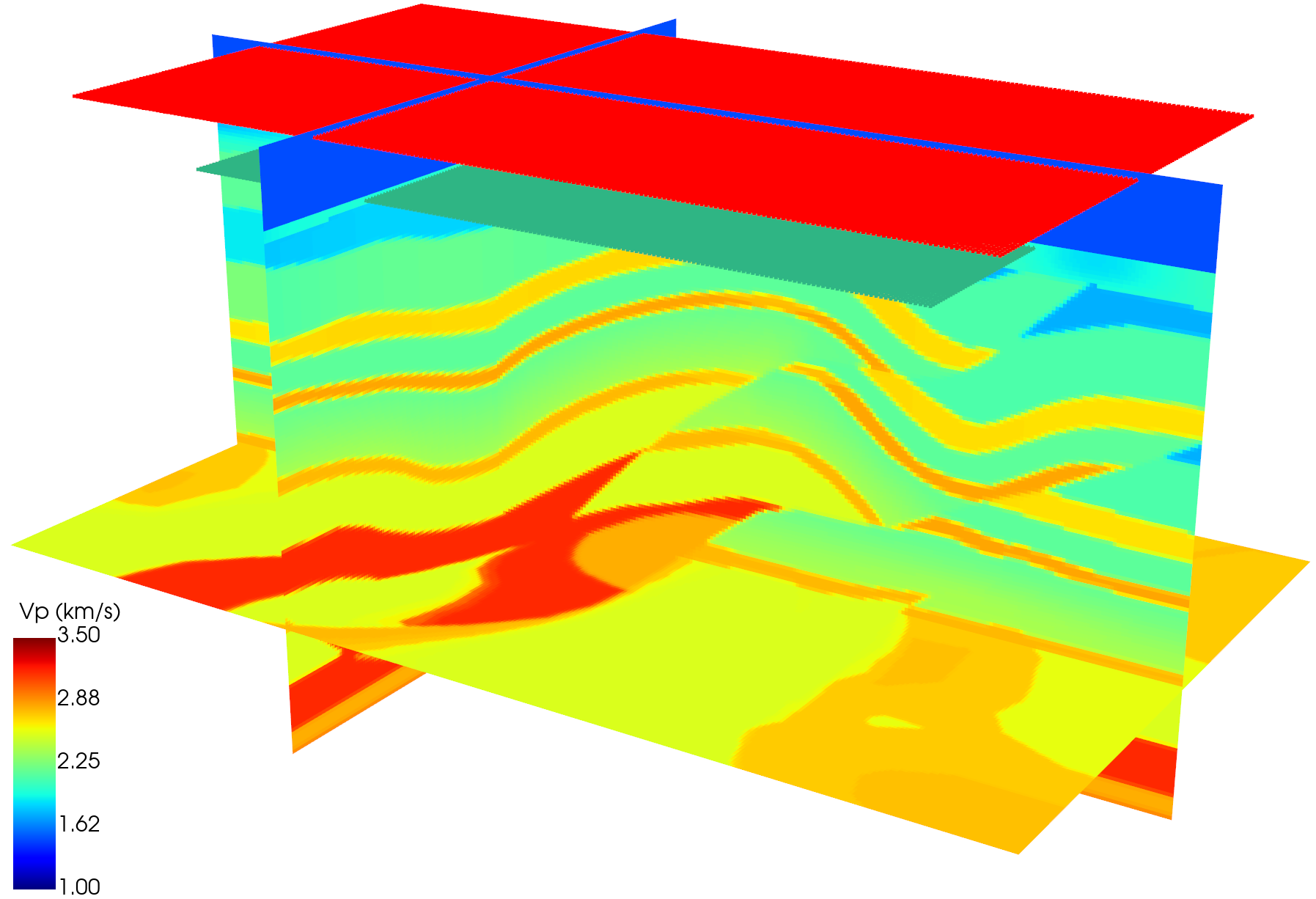}\label{fig:3dvel}}
     \subfloat[][]{\includegraphics[width=.4\linewidth]{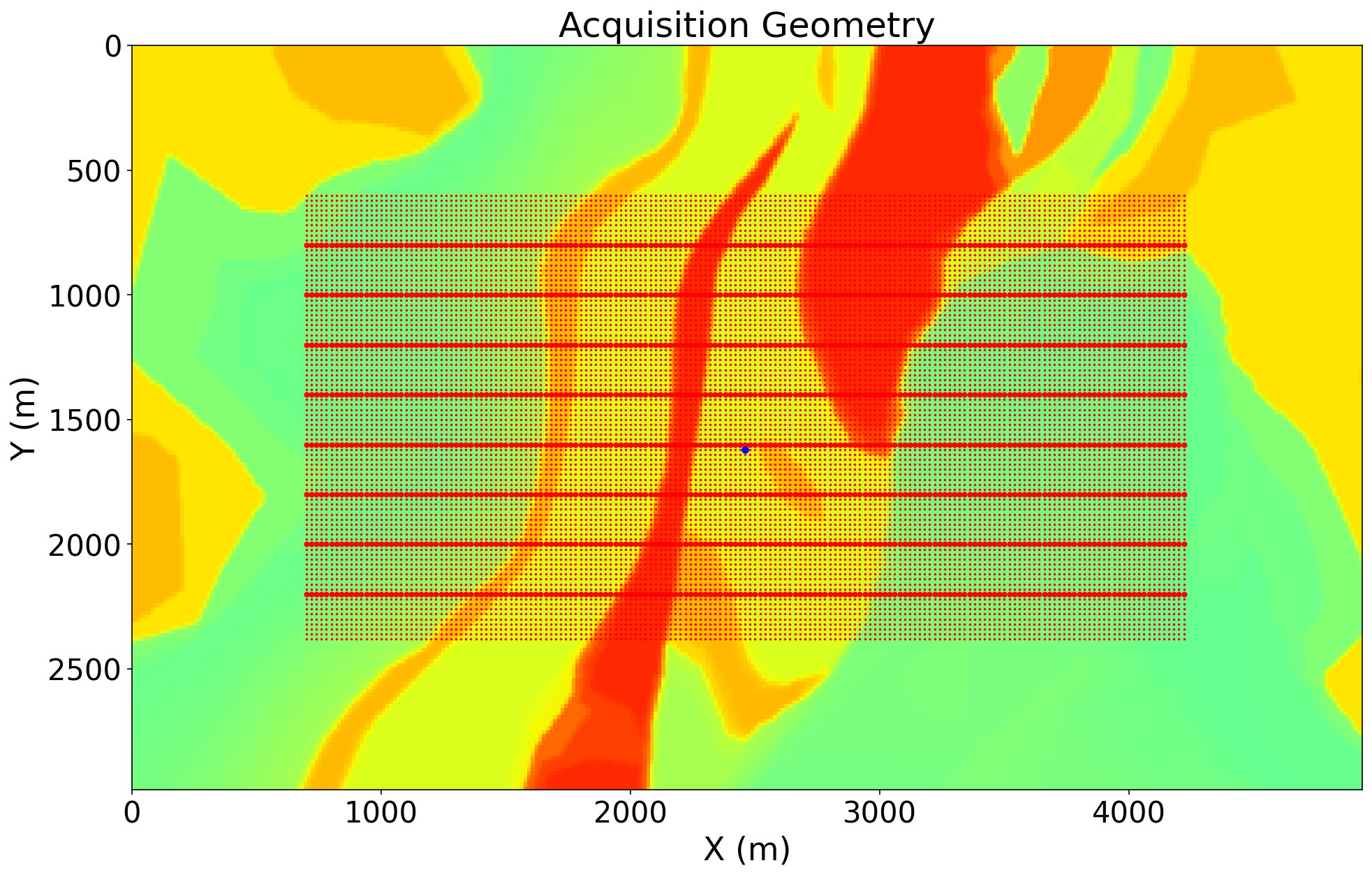}\label{fig:gmt}}
     \caption{(a) The 3D EAGE/SEG Overthrust model used to generate the down- and up-going wavefield and (b) The receiver layout, with blue dots representing the virtual source and bold lines indicating the receiver lines, from which the reconstructed wavefield by the frequency-domain MDD is extracted for comparison in time-offset domain.}
     \label{fig:velandgmt}
\end{figure}

To evaluate the effectiveness of the proposed algorithm, we design a large-scale 3D synthetic example based on the 3D EAGE/SEG Overthrust model, as shown in Figure \ref{fig:3dvel}, with the geometry illustrated in Figure \ref{fig:gmt}. The setup includes 177 receiver lines and 217 source lines. Each receiver line is equipped with 90 receivers, and each source line contains 120 sources. Both sources and receivers are placed at intervals of $\SI{20}{\meter}$ in both directions (i.e., along and across the lines).
In Figure \ref{fig:gmt}, we display all receivers as red dots while omitting the shot locations because the original shot information, which indexes the down- and up-going wavefields, is eliminated after applying MDD. Additionally, we use eight red lines and a blue dot in Figure \ref{fig:gmt} to highlight the receiver lines and the virtual source, respectively. These elements mark the positions where the reconstructed wavefield will be extracted for visualization purposes, allowing us to compare the results obtained from different MDD solvers.

With the geometry and velocity model prepared, pressure and particle velocity data are generated using a \SI{25}{\hertz} wavelet over a duration of 4.5 seconds. This is achieved via the finite difference method, with data sampled at 4 ms intervals. Wavefield separation is then performed to extract the down- and up-going pressure wavefield components, which are subsequently transformed into the frequency domain.
The resulting dataset, after transformation into the frequency domain, consists of two stacks of $N=200$ matrices each: one for $\mathbf{D}$ and one for $\mathbf{U}$. Both matrices have dimensions of $26040 \times 15930$, where $26040=217 \times 120$ and $15930=177 \times 90$. The unknown Green's function matrices $\mathbf{X}$ have dimensions of $15930 \times 15930$. For each frequency, storing the elements of the three dense matrices $\mathbf{D}$, $\mathbf{U}$, and $\mathbf{X}$ requires approximately 8 gigabytes of memory. Further details on dataset creation and pre-processing can be found in \cite{hongravasi2023} and
\cite{Ltaief2023}.

\begin{figure}
     \centering
     \subfloat[][]{\includegraphics[width=.8\linewidth]{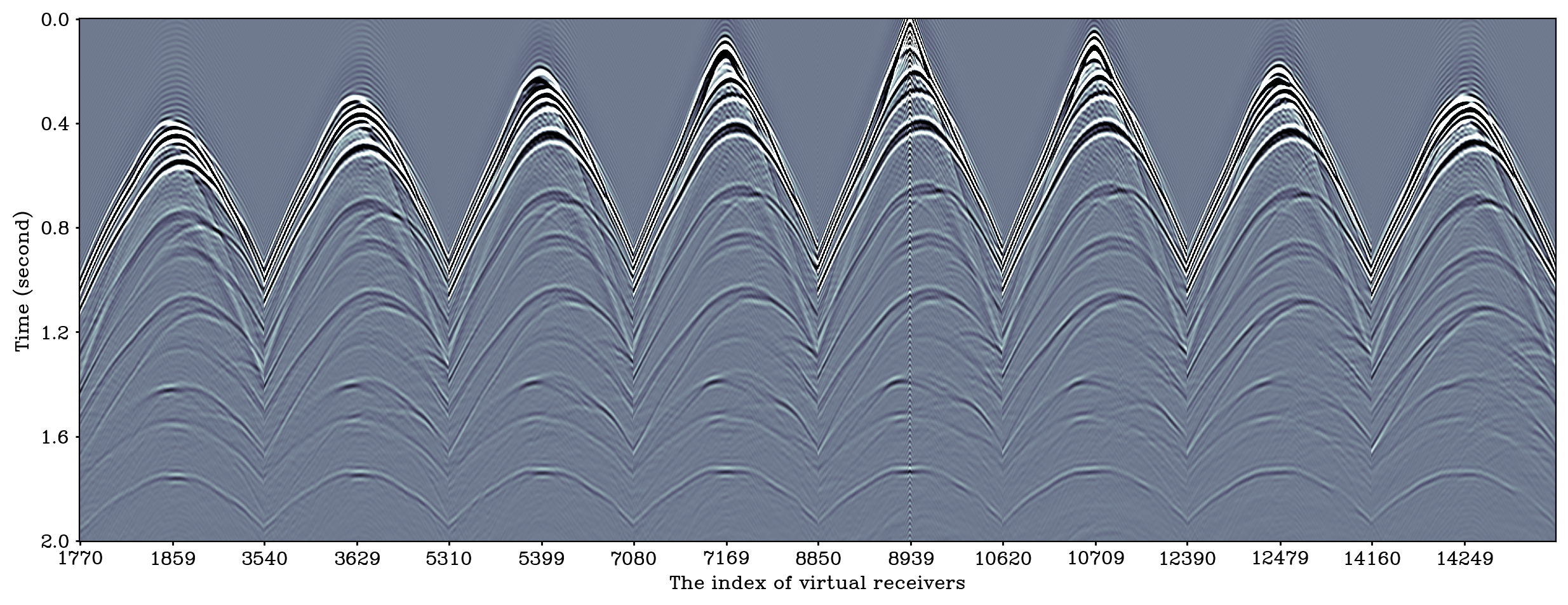}\label{fig:t_x_tr}}
     \hspace{0.5cm}
          \subfloat[][]{\includegraphics[width=.8\linewidth]{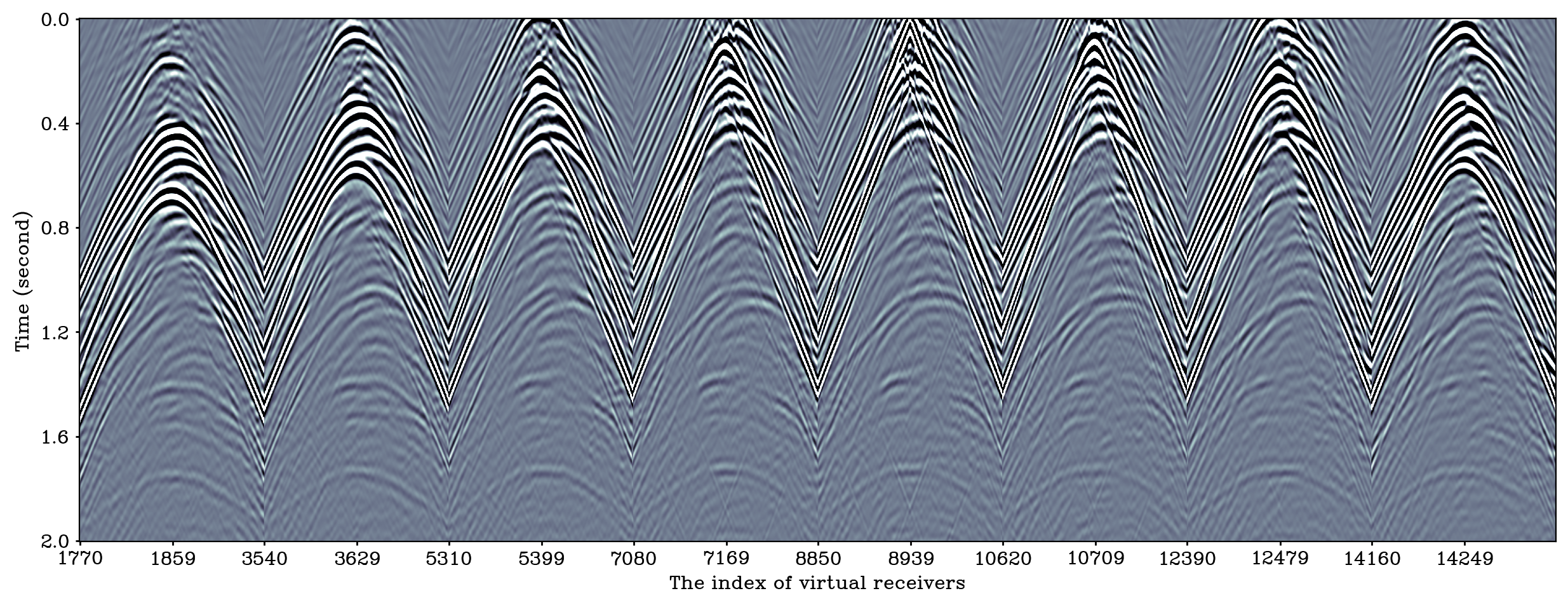}\label{fig:t_x_adj}}
     \hspace{0.5cm}
     \subfloat[][]{\includegraphics[width=.8\linewidth]{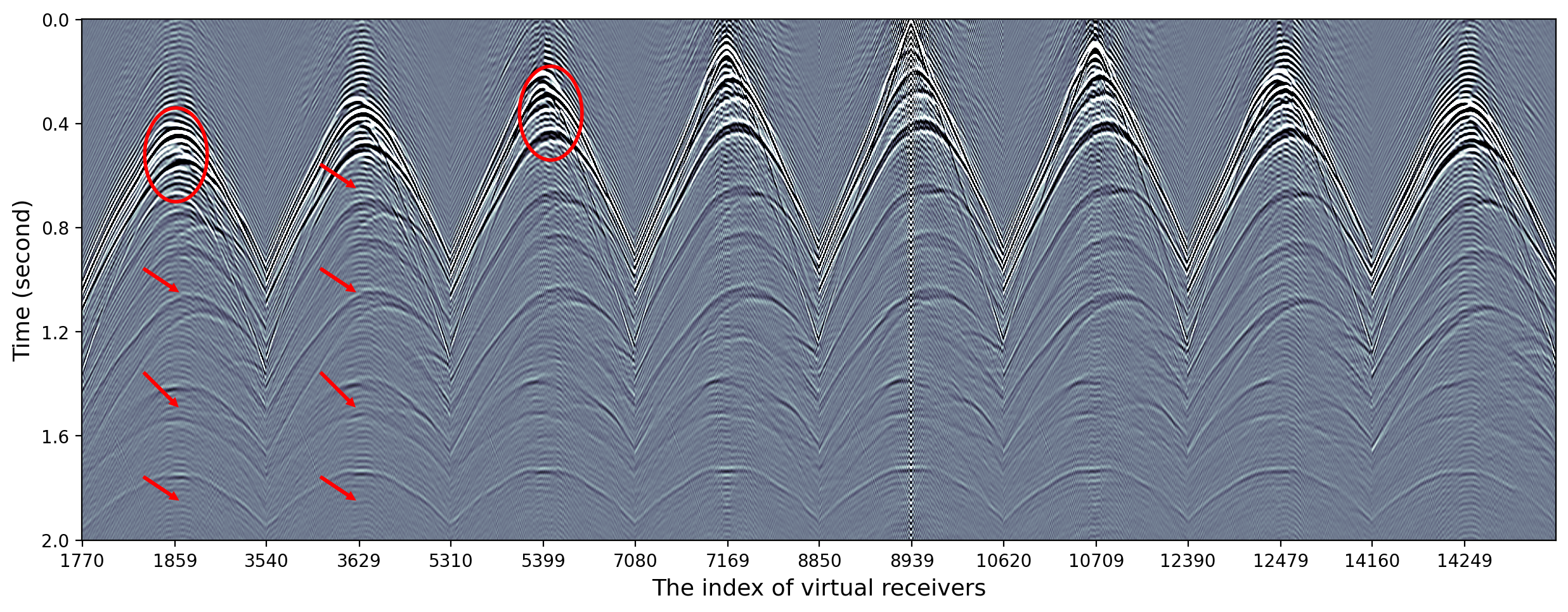}\label{fig:t_x_bmk}}
     \caption{(a) The true  Green's function in the time-offset domain and the retrieved Green's function with (b) the cross-correlation methodand (c) the benchmark solver.}
     \label{x1_vs_x2}
\end{figure}
\begin{figure}
     \centering
     \subfloat[][]{\includegraphics[width=.4\linewidth]{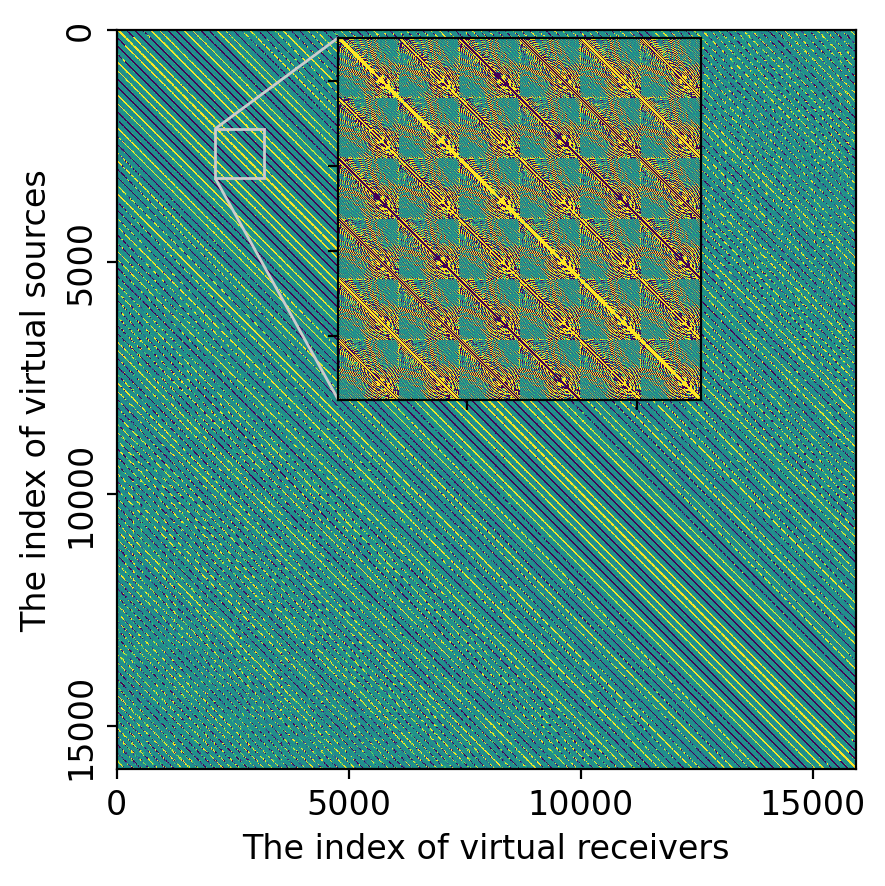}\label{fig:x_no_sort}}
     \subfloat[][]{\includegraphics[width=.4\linewidth]{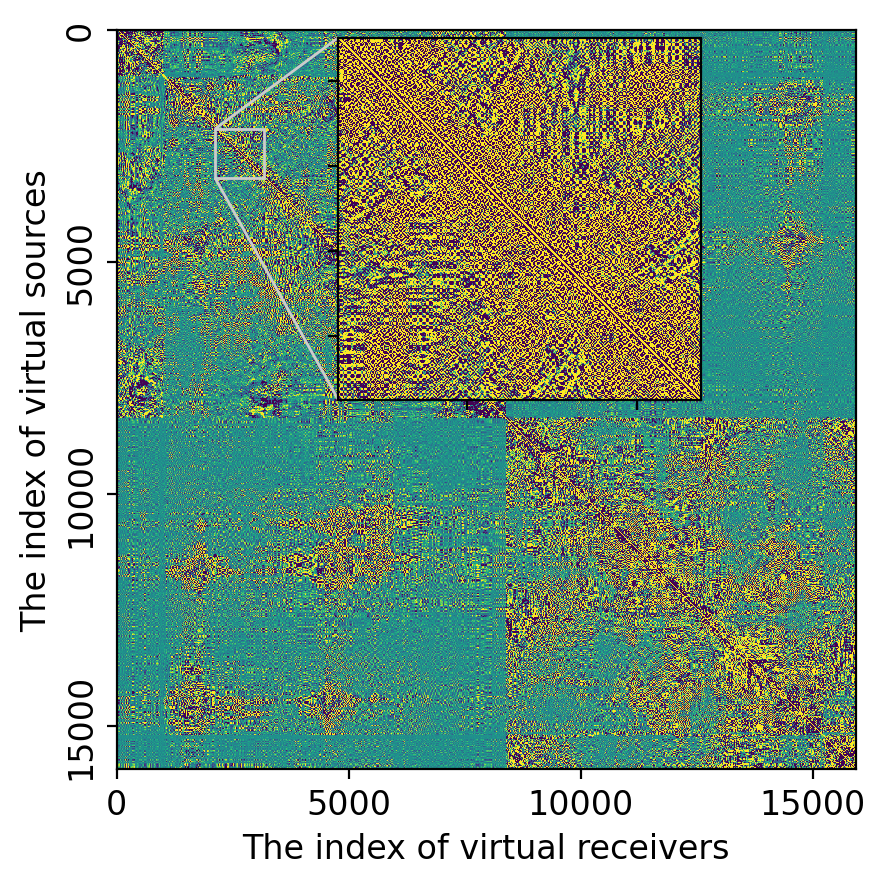}\label{fig:x_sort}}
     \caption{The reference Green's function corresponding to the dominant frequency with (a) natural ordering and (b) re-ordering following the Hilbert space-filling curve.}
     \label{fig:x1_vs_x2}
\end{figure}

\begin{figure}
     \centering
     \subfloat[][]{\includegraphics[width=.6\linewidth]{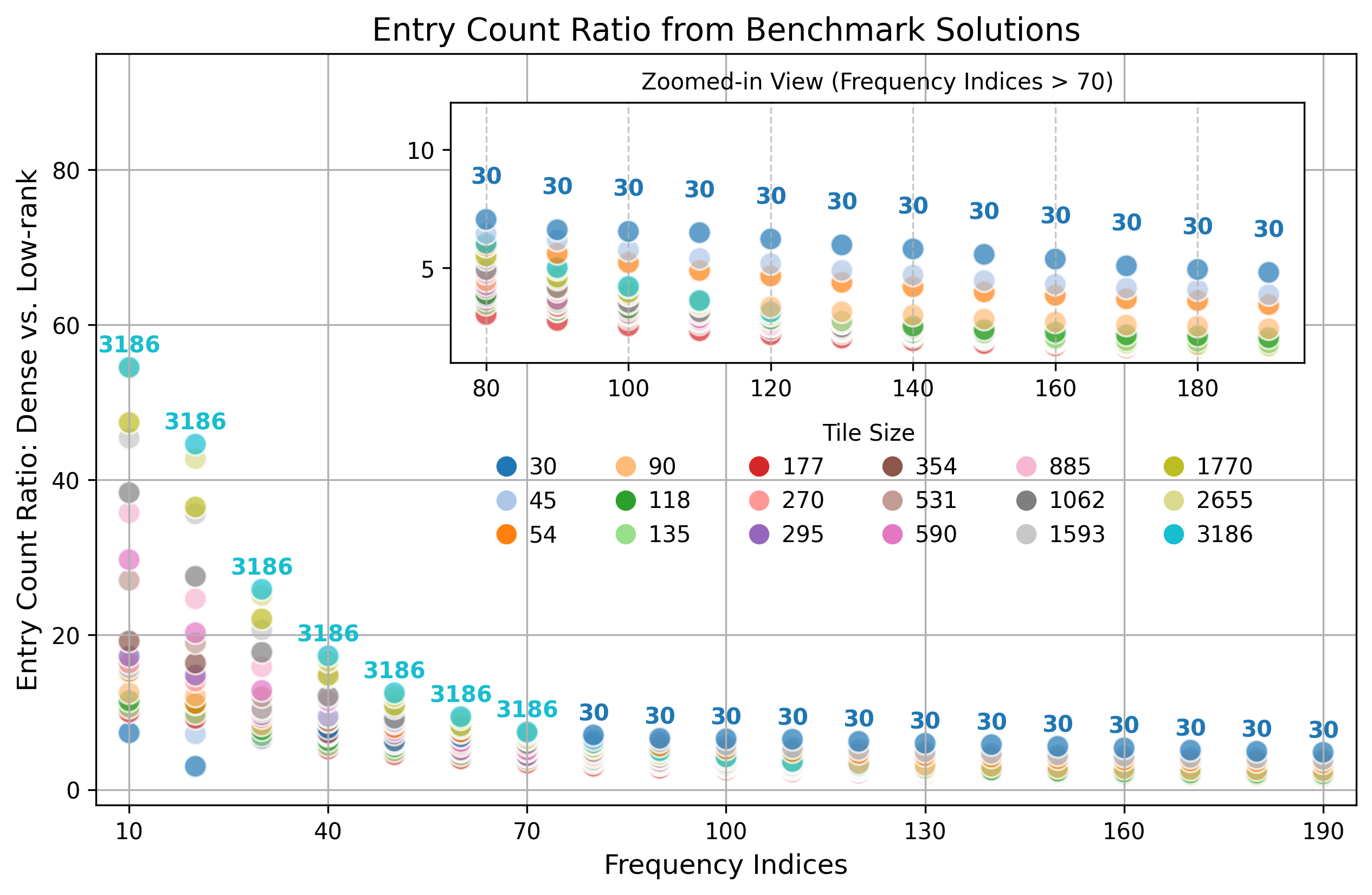}\label{fig:entryratio1}}
     \hspace{0.5cm}
     \subfloat[][]{\includegraphics[width=.6\linewidth]{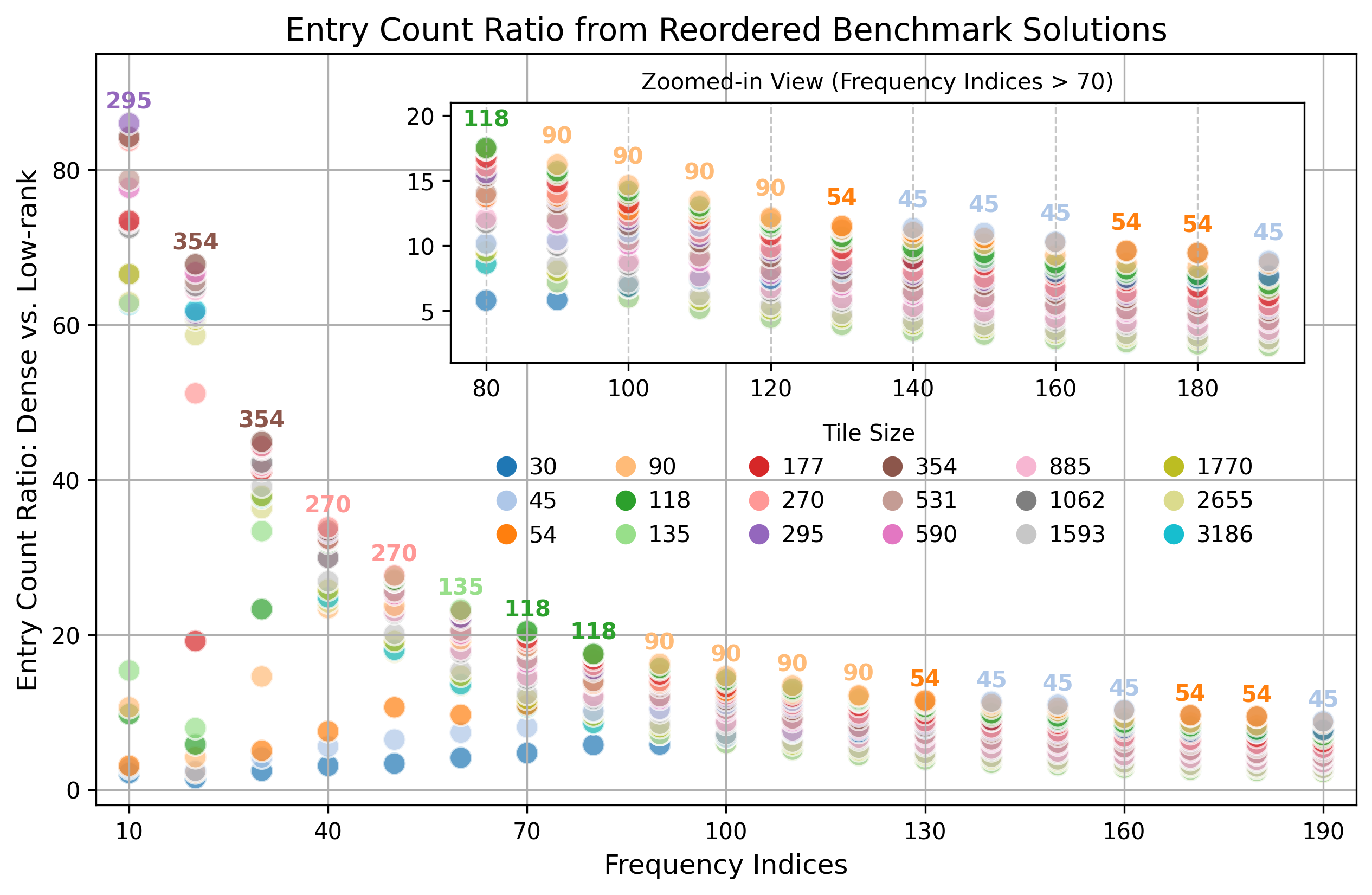}\label{fig:entryratio2}}
     \caption{The entry count ratio analyzed based on the solution from the benchmark solver: (a) using natural ordering and (b) using ordering based on the Hilbert curve.}
     \label{fig:entryratio}
\end{figure}
\subsection{MDD Solver Evaluation}
Before applying any MDD solver, we generate the true solution using the velocity model shown in Figure~\ref{fig:3dvel}. The corresponding time-offset domain result is displayed in Figure~\ref{fig:t_x_tr}. Next, we compute the adjoint solution for multi-dimensional cross-correlation, which does not require inversion. Each frequency-domain solution is obtained using $ \mathbf{X}_{\mathrm{adj}} = \mathbf{D}^\mathrm{H} \mathbf{U}$.  
The resulting time-offset domain adjoint solution is shown in Figure~\ref{fig:t_x_adj}, which reveals not only its low temporal resolution but also that not all surface-related multiples are removed. 
Figure~\ref{fig:t_x_bmk} shows the reconstructed Green's function using a benchmark MDD solver, specifically the least-squares preconditioned solver described in equation~\ref{eq:bench_solver}. This serves as a reference against which we evaluate the accuracy of the proposed MDD solver.
Compared to the true solution shown in Figure~\ref{fig:t_x_tr}, the inversion result obtained using the benchmark solver demonstrates significantly better accuracy than the adjoint result, which is merely the cross-correlation of the down-going and up-going wavefields. This highlights the necessity of using MDD, rather than relying solely on the adjoint operator, to accurately reconstruct the unknown Green's function.

Since the effectiveness of low-rank factorization for an inverse problem depends on the degree of low-rankness of the unknown matrix, it is essential to demonstrate that reordering the rows and columns of the solution matrix improves its low-rank compressibility. To illustrate this, we use the Hilbert space-filling curve-based sorting algorithm. Figure~\ref{fig:x_no_sort} displays the reference solution at the dominant frequency of \SI{25}{\hertz}, where receivers and virtual sources are arranged according to the cable-line ordering. As shown in the zoomed section, this ordering naturally results in a tiled structure for the Green's function in the frequency domain, with each tile measuring $177\times177$. Notably, each tile demonstrates diagonal dominance, leading to a high numerical rank. In contrast, Figure~\ref{fig:x_sort} presents the same matrix after reordering its rows and columns based on the Hilbert curve computed over the acquisition geometry. This reordering largely eliminates the tiled structure and concentrates most of the Green's function energy along the main diagonal.

We then use different tile sizes, as listed in Figure~\ref{fig:entryratio}. These tile sizes are determined based on the factorization of the total number of receivers into prime factors, $15930 = 2 \cdot 3^3 \cdot 5 \cdot 59$, to test the compression ratio. The process involves the following steps:  

\begin{enumerate}
    \item For a given frequency and tile size, we divide the MDD solution obtained from the benchmark solver into tiled blocks.
    \item We perform SVD on each tile for the specified tile size.
    \item For each tile, we determine its numerical rank $k$ based on the maximum singular value $v_0$ among all tiles. Specifically, $k = \mathrm{argmax}(v < \epsilon v_0)$, where $v$ represents the singular values of a tile and $\epsilon = 0.01$ in this study. If no singular value satisfies the condition $v < \epsilon v_0$, the tile is considered numerically full rank.
\end{enumerate}

To identify the optimal tile size for a given reference solution at a specific frequency, we calculate the total number of entries across all factors and divide $15930^2$ (the total number of parameters in the dense matrix $\mathbf{X}$) by this number. We refer to this as the \textit{entry count ratio}.  
Figure~\ref{fig:entryratio1} presents the entry count ratio results when the reference solution is ordered according to cable lines, with the optimal tile size indicated for each frequency. In contrast, Figure~\ref{fig:entryratio2} shows the results when the reference solution is reordered using the Hilbert space-filling curve based on receiver geometry, along with the corresponding optimal tile sizes for each frequency.  
In both cases, we observe that the entry count ratio decreases with increasing frequency, indicating that the total numerical rank increases as the frequency rises. However, comparing Figures~\ref{fig:entryratio1} and \ref{fig:entryratio2}, it is evident that reordering significantly improves the low-rank property. For instance:
\begin{itemize}
    \item At frequency index 10, the entry count ratio with cable-line ordering is below 60 (optimal tile size: 3186), while with Hilbert ordering, it exceeds 80 (optimal tile size: 295).
    \item At frequency index 190, the entry count ratio with cable-line ordering is around 5 (optimal tile size: 30), while with Hilbert ordering, it is approximately 10 (optimal tile size: 45).
\end{itemize}
\begin{figure}
     \centering
     \subfloat[][]{\includegraphics[width=.4\linewidth]{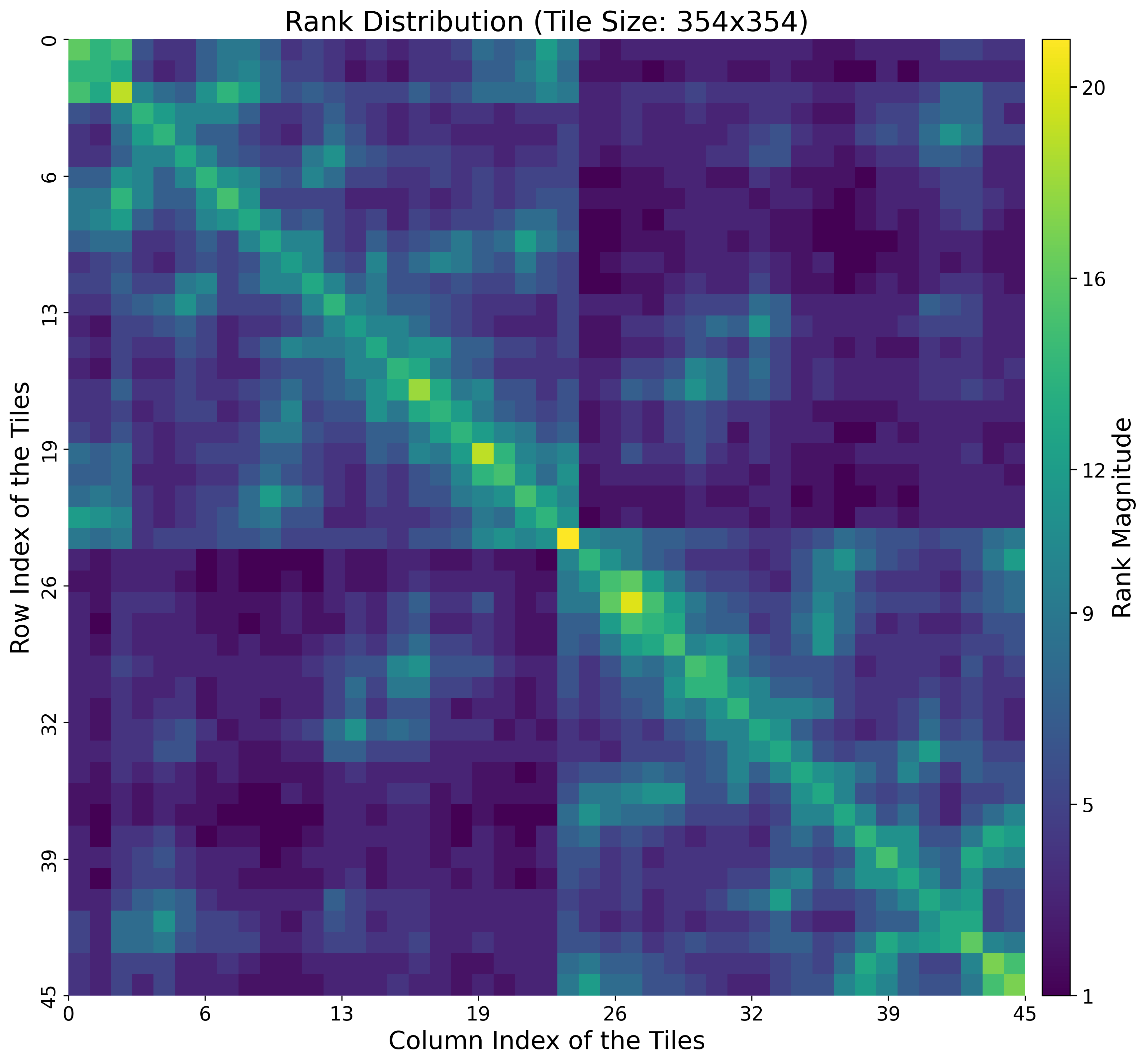}\label{fig:rankdist1}}
     \subfloat[][]{\includegraphics[width=.4\linewidth]{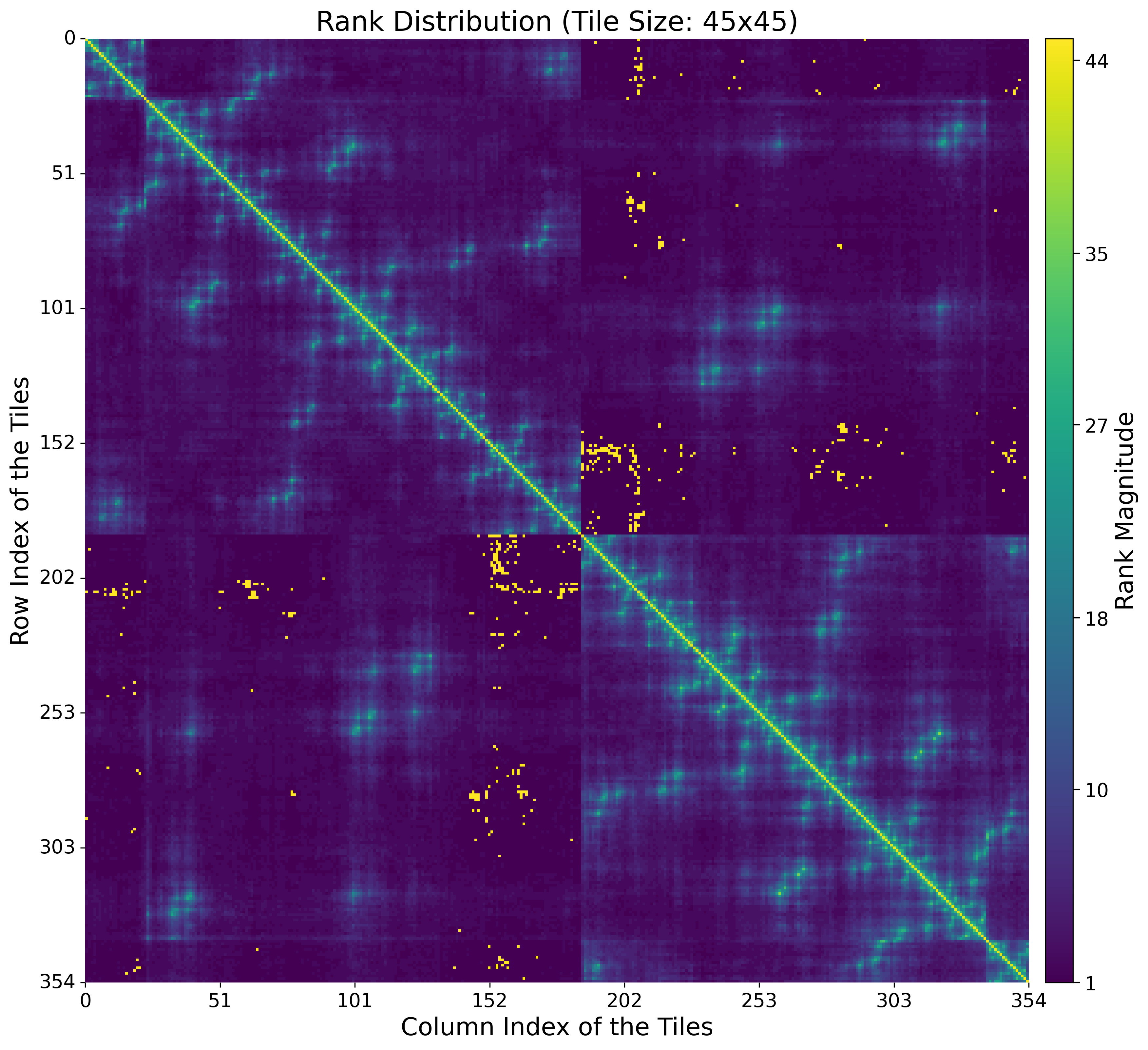}\label{fig:rankdist2}}
     \caption{The distribution of the local numerical rank of the benchmark solver's solution corresponding to the frequency index: (a) at 20 and (b) at 160.}
     \label{rank_dist}
\end{figure}
\begin{figure}
     \centering
     \subfloat[][]{\includegraphics[width=.9\linewidth]{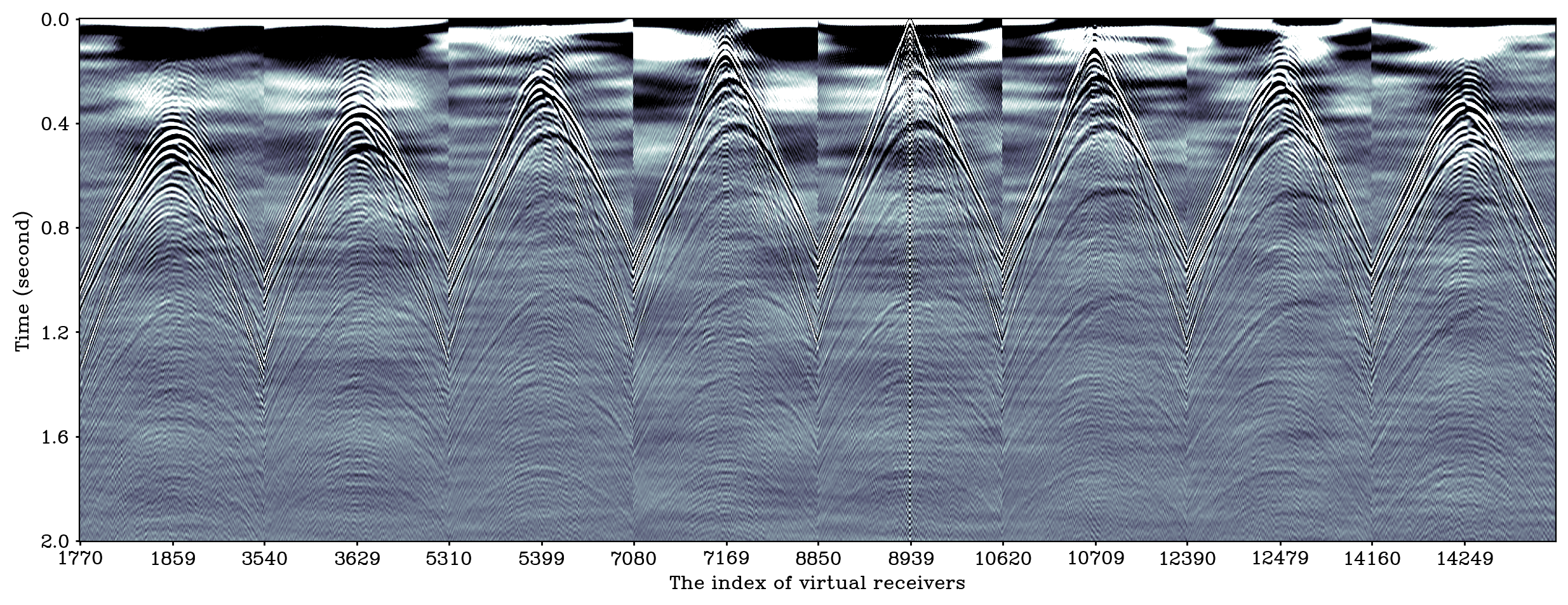}\label{fig:t_x_nosort_r3}}
     \hspace{0.5cm}
     \subfloat[][]{\includegraphics[width=.9\linewidth]{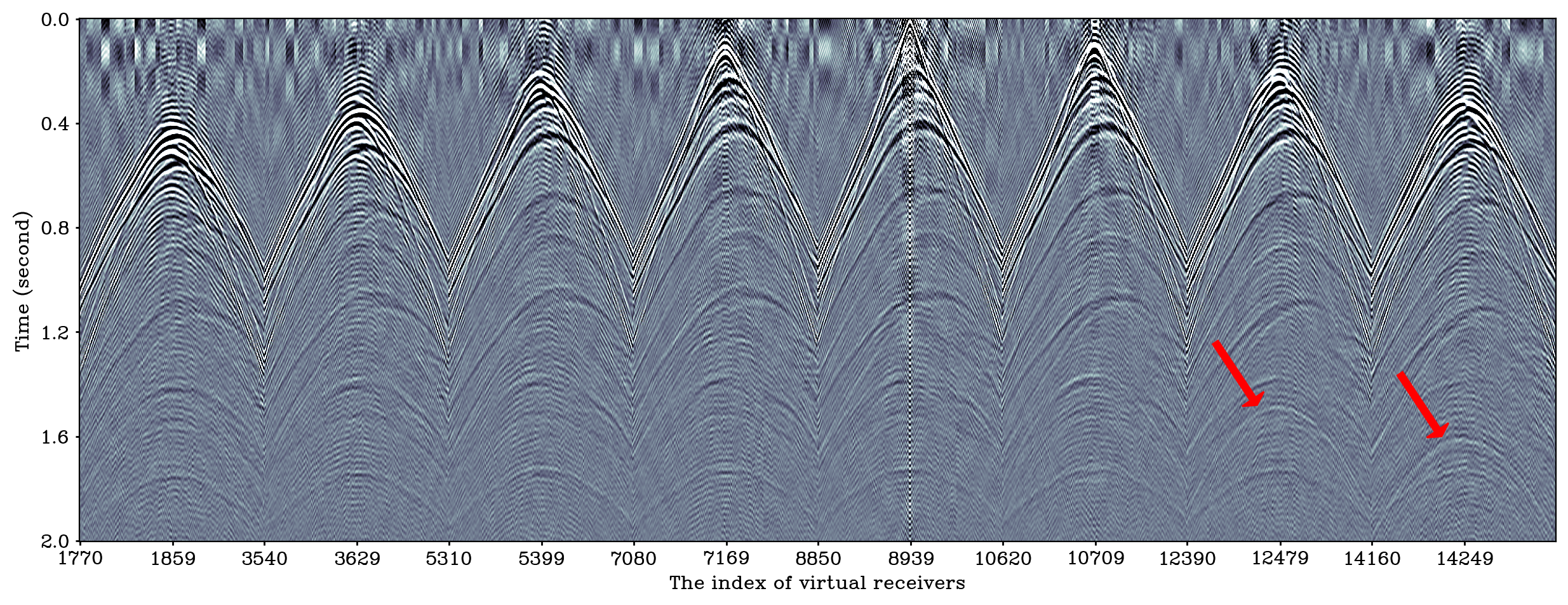}\label{fig:t_x_rsort_r3}}
     \hspace{0.5cm}
     \subfloat[][]{\includegraphics[width=.9\linewidth]{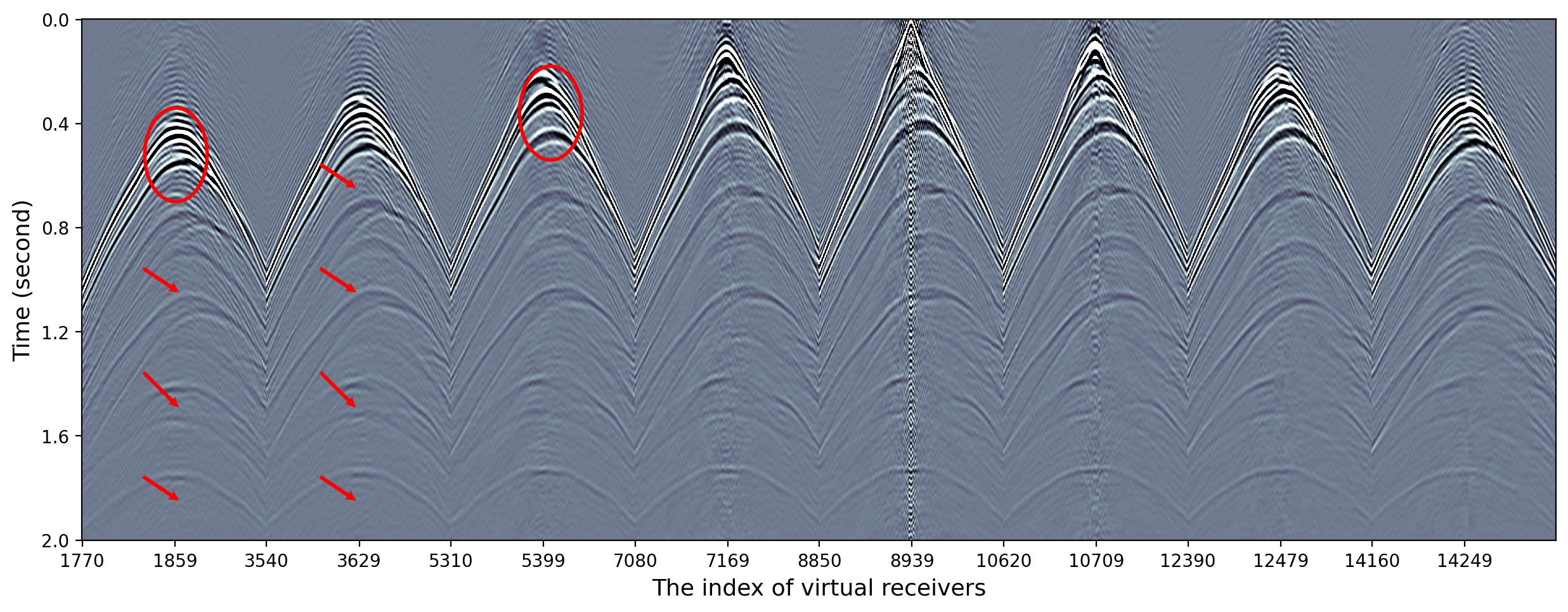}\label{fig:t_x_tile64_20diags_r3}}
     \caption{The time-domain Green's function reconstructed by global low-rank approximation with (a) the row-major re-ordering, (b) re-ordering based on the Hilbert curve, and (c) the proposed tile low-rank approximation.}
     \label{mdd_tlr_tx}
\end{figure}

We now examine the rank distribution with adaptive tile sizes. Figure~\ref{fig:rankdist1} illustrates the rank distribution for a frequency index of 20 (corresponding to a frequency of $\SI{5.6}{\hertz}$), representing the rank distribution for low frequencies. For this frequency index, with an optimal tile size of 354, we observe that all tiles are rank-deficient, and many off-diagonal tiles exhibit very low ranks. Even the diagonal tiles, while higher in rank, remain significantly rank-deficient, with their maximum rank being quite low compared to the tile size of 354.
Figure~\ref{fig:rankdist2} presents the rank distribution for a frequency index of 160 (corresponding to a frequency of $\SI{44.4}{\hertz}$), representing the rank distribution for high frequencies. For this frequency index, with an optimal tile size of 45, most tiles are still rank-deficient, and a large portion of the off-diagonal tiles maintain very low ranks relative to the tile size. However, the rank distribution at this higher frequency reveals some full-rank tiles, particularly among the diagonal tiles and a few off-diagonal tiles.
This observation highlights the advantage of adaptively representing the unknown matrix with varying tile sizes across different frequencies.

Before applying the proposed solver based on adaptive tile low-rank factorization, we first attempt to solve the 3D MDD problem using a relatively straightforward approach, namely the reciprocity-aware global low-rank factorization \citep[]{chen2023physicsaware}, which has been successful for 2D MDD problems. This experiment aims to determine whether this method can produce satisfactory results for 3D problems. The numerical results are shown in Figure~\ref{fig:t_x_nosort_r3}, where the unknown Green's function is approximated as $\mathbf{X} = \mathbf{Q}\mathbf{Q}^\mathbf{T}$, with $\mathbf{Q}$ having dimensions of $15930 \times 5310$. As observed, the resulting solution is unsatisfactory compared to the benchmark solver.
To investigate whether reordering the down- and up-going wavefields based on the Hilbert space-filling curve (i.e., first obtaining $\mathbf{X}_h$ and then converting it to $\mathbf{X}$) improves low-rank compressibility and the overall solution quality, we applied this operation. The results are presented in Figure~\ref{fig:t_x_rsort_r3}. While both the natural sorting and Hilbert curve-based reordering approaches achieve the same level of memory savings, the Hilbert curve reordering effectively eliminates the low-frequency noise observed in the natural sorting approach (Figure~\ref{fig:t_x_nosort_r3}).
However, when compared to the benchmark MDD solution in Figure~\ref{fig:t_x_bmk}, both results are of lower quality. Noise is introduced into the solution, and, more importantly, remnant free-surface multiples remain present in the Green's function, as indicated by the red arrows in Figure~\ref{fig:t_x_rsort_r3}.

In the following, we demonstrate how the proposed MDD solver not only resolves the aforementioned issues but also delivers superior results compared to the benchmark solver. In this approach, the unknown Green's function, $\mathbf{X}$, is adaptively divided into tiles, allowing the matrix corresponding to different frequencies to be partitioned with varying tile sizes. The optimal tile size and numerical rank for each tile are determined based on the solution obtained from the benchmark solver. Each tile is then represented using either a factorization-based low-rank approximation or a preconditioned dense form, depending on its numerical rank.
Figure~\ref{fig:t_x_tile64_20diags_r3} shows the MDD result obtained using the proposed solver. Unlike the global low-rank approximation with Hilbert curve-based reordering, our reciprocity-aware tile low-rank parameterization, combined with Hilbert curve reordering, achieves a significantly higher signal-to-noise ratio and eliminates the striped artifacts observed in the solution from the global low-rank factorization with reordering. Furthermore, the inverted Green's function computed using the proposed MDD solver is free of remnant multiples.
When comparing the results of the proposed method to those of the benchmark MDD solver, two significant improvements are observed when using the true solution as a reference. These improvements are highlighted in both solutions using red circles and red arrows. First, the proposed method demonstrates superior de-multiple performance, as shown by the regions marked with red circles. Second, the artifacts present in the near-offset regions of the benchmark solution, as indicated by the red arrows, are completely eliminated in the solution obtained with the proposed method.

\begin{figure}
     \centering
\includegraphics[width=.7\linewidth]{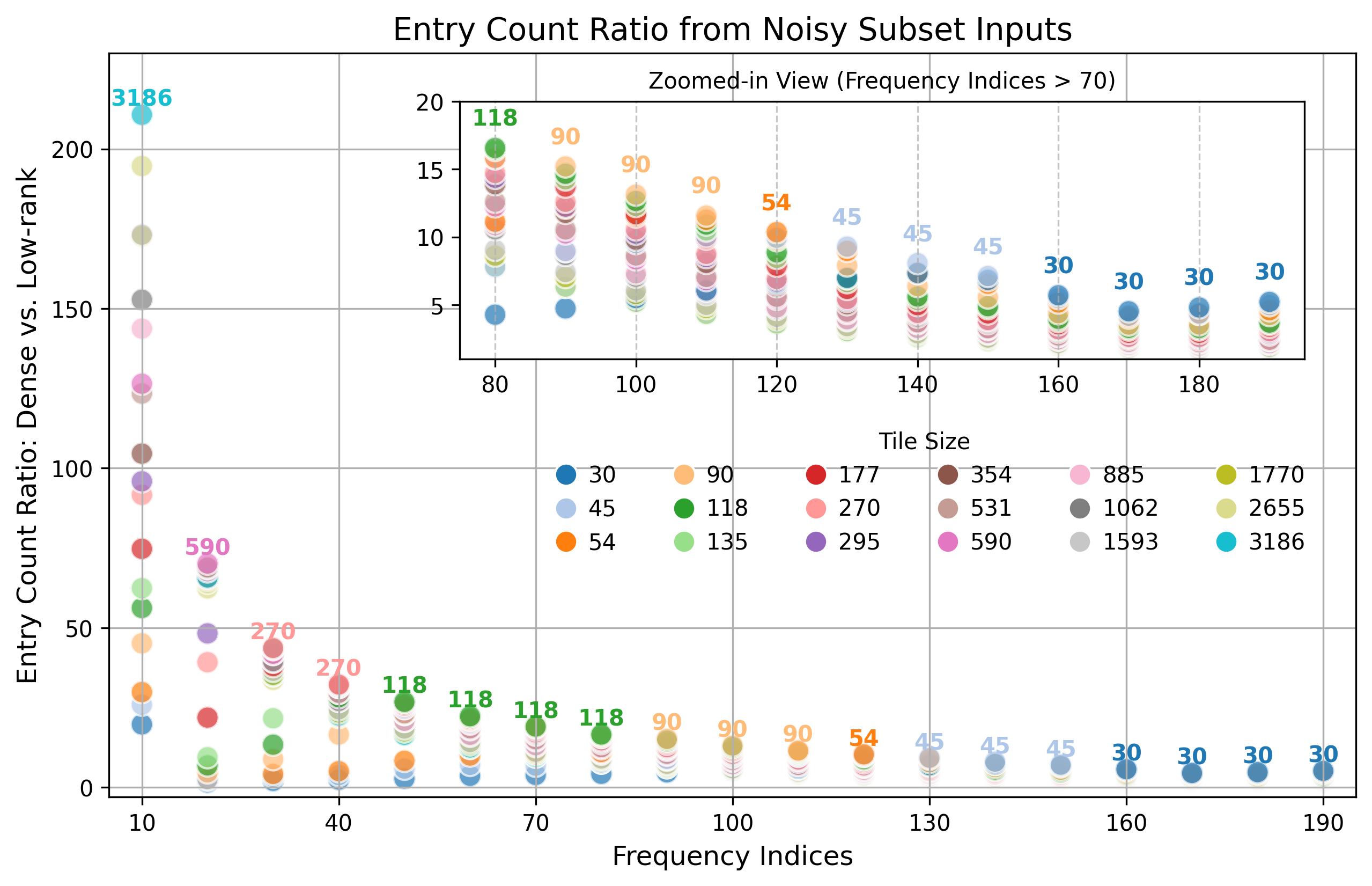}
     \caption{The entry count ratio analyzed based on the solution from the benchmark solver using ordering based on the Hilbert curve. In this scenario, 50$\%$ of the total 26,040 shots are selected, with random noise added to both the down-going and up-going wavefields.}
     \label{fig:entryratio50}
\end{figure}
\begin{figure}
     \centering
     \subfloat[][]{\includegraphics[width=.9\linewidth]{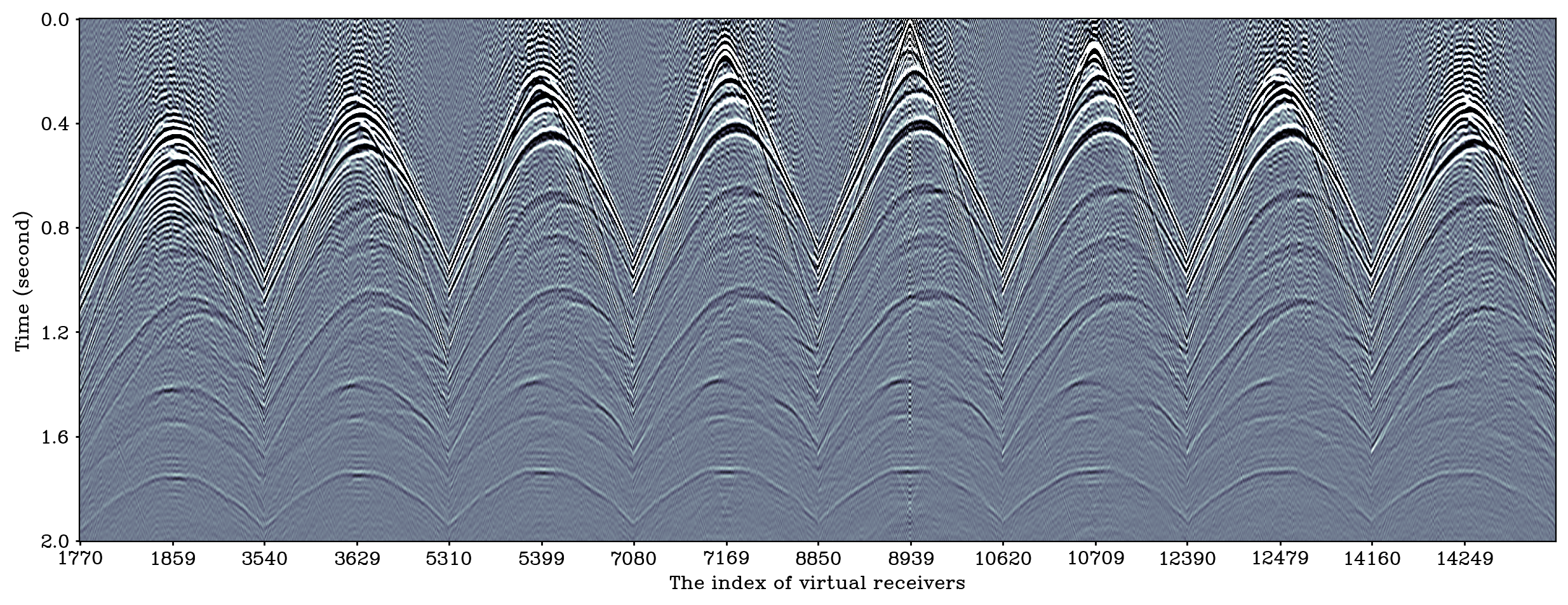}\label{fig:t_x_tik_perc50}}
     \hspace{0.5cm}
     \subfloat[][]{\includegraphics[width=.9\linewidth]{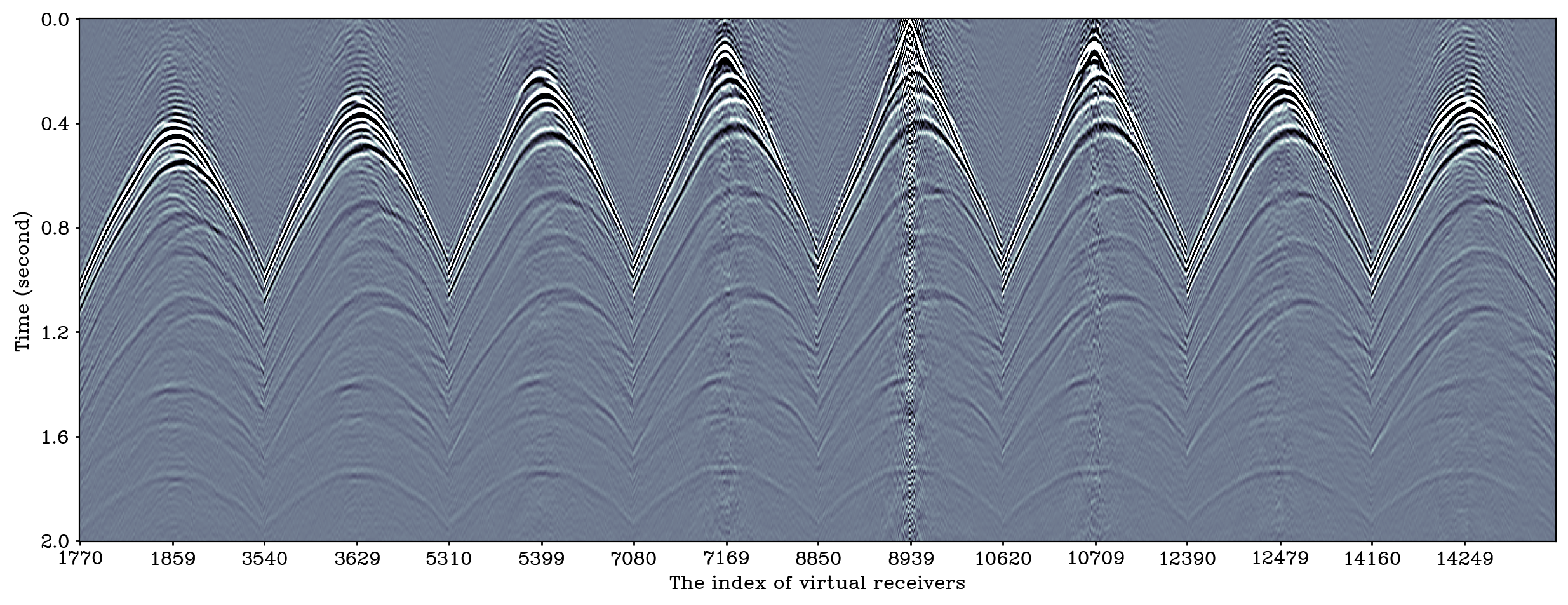}\label{fig:t_x_tlr_perc50}}
     \caption{The time-domain Green's function estimated by (a) the benchmark solver and (b) the proposed MDD solver with the noisy subset input. }
     \label{mdd_tlr_tik_perc50}
\end{figure}
\begin{figure}
     \centering
\includegraphics[width=.7\linewidth]{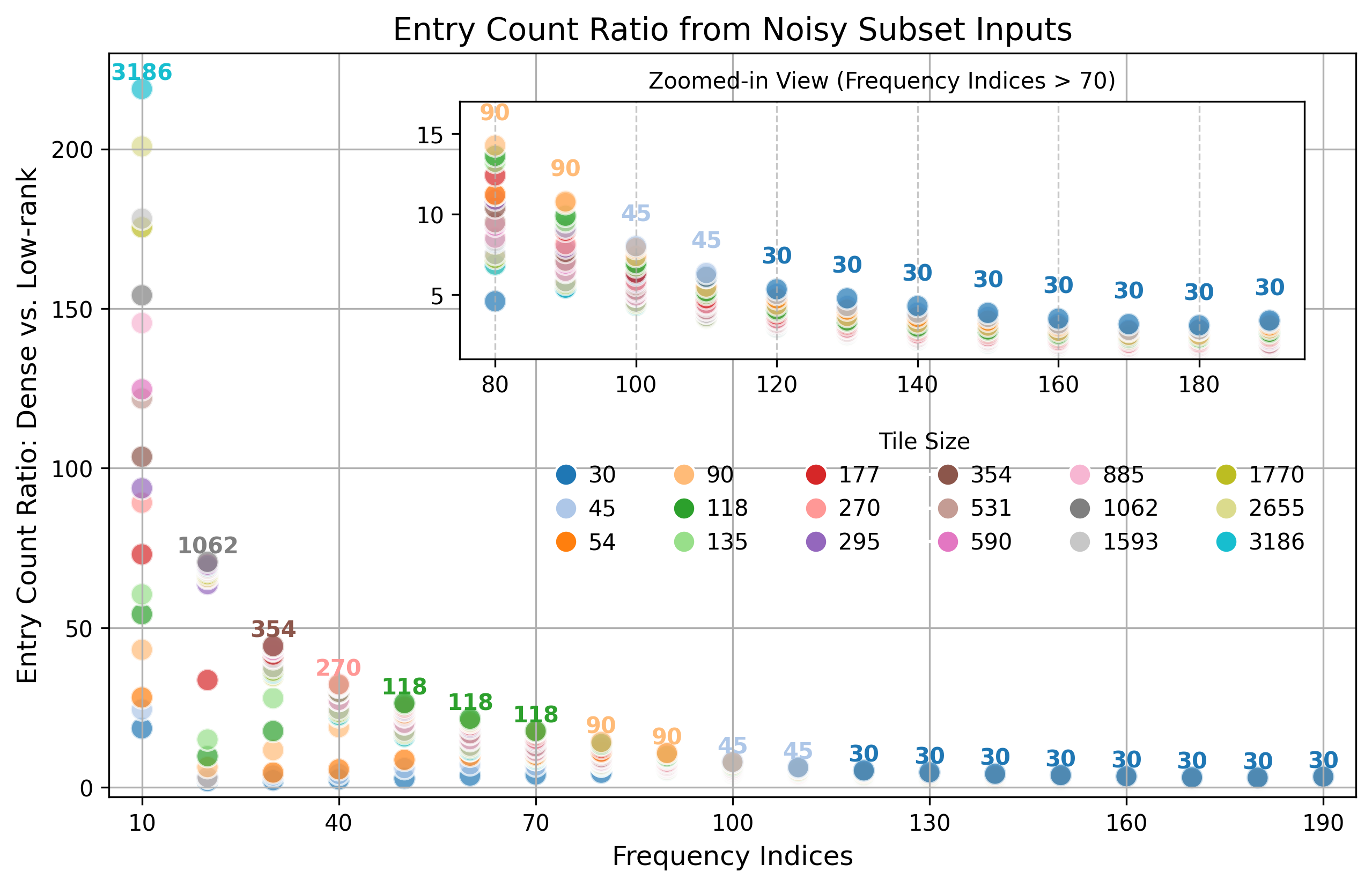}
     \caption{The entry count ratio analyzed based on the solution from the benchmark solver using ordering based on the Hilbert curve. In this scenario, 25$\%$ of the total 26,040 shots are selected, with random noise added to both the down-going and up-going wavefields.}
     \label{fig:entryratio25}
\end{figure}
\begin{figure}
     \centering
     \subfloat[][]{\includegraphics[width=.9\linewidth]{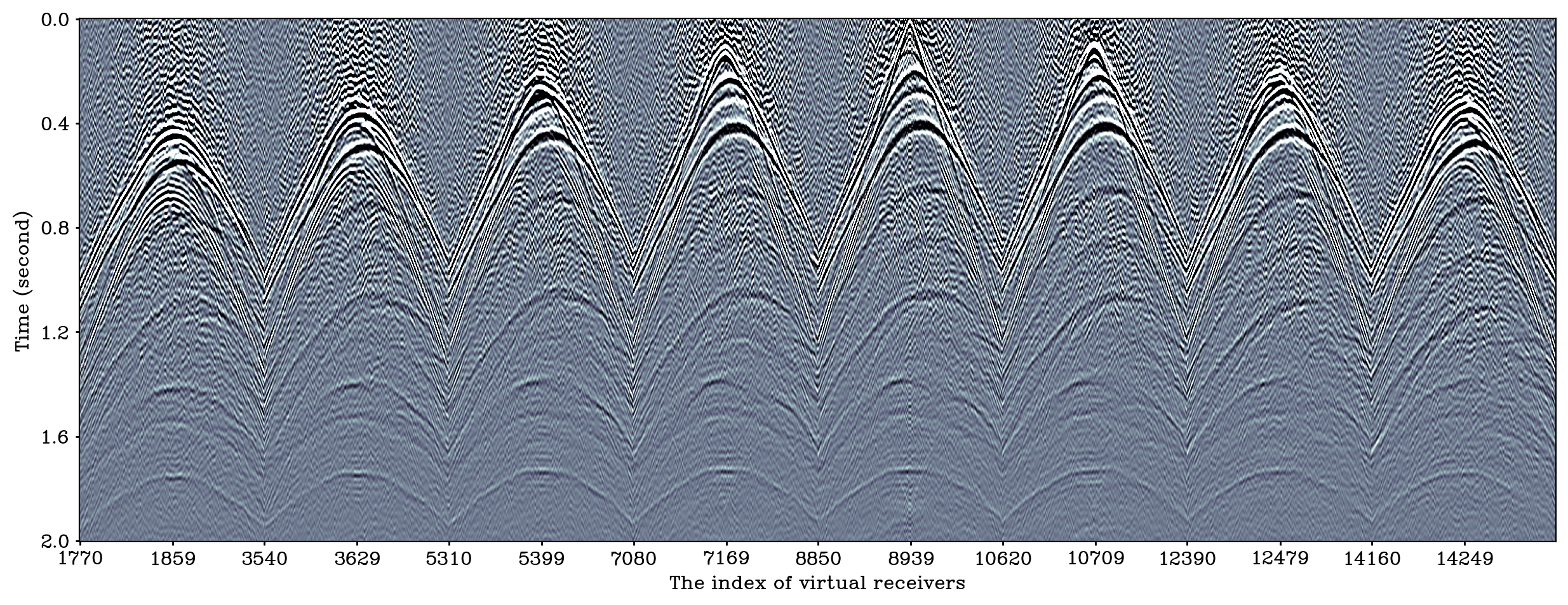}\label{fig:t_x_tik_perc25}}
     \hspace{0.5cm}
     \subfloat[][]{\includegraphics[width=.9\linewidth]{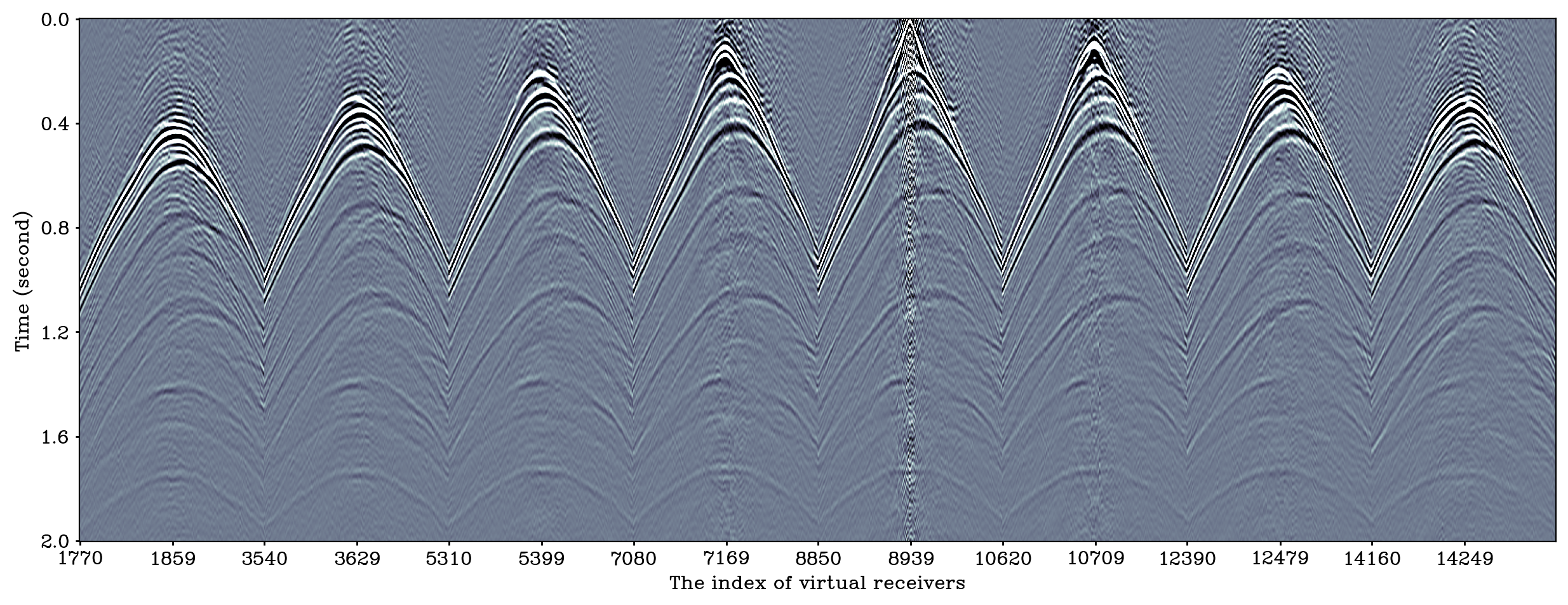}\label{fig:t_x_tlr_perc25}}
     \caption{The time-domain Green's function estimated by (a) the benchmark solver and (b) the proposed MDD solver with the noisy subset input. }
     \label{mdd_tlr_tik_perc25}
\end{figure}
\subsection{Noise and Sparse Sampling Case}
To further evaluate the robustness of the proposed MDD solver compared to the benchmark solver, we randomly select 50$\%$ of the shots (i.e., half of the rows in $\mathbf{D}$ and $\mathbf{U}$ are randomly discarded) and add Gaussian noise to both the down-going and up-going wavefields in the time-offset domain. As before, the first step involves estimating the optimal tile size and numerical rank for each tile at each frequency, using the solution from the benchmark solver, as shown in Figure~\ref{fig:t_x_tik_perc50}. The same strategy is applied to determine the optimal tile size and numerical ranks, and the results are presented in Figure~\ref{fig:entryratio50}.

When comparing the entry count ratio in Figure~\ref{fig:entryratio50} to that in Figure~\ref{fig:entryratio2}—which is obtained from analyzing the benchmark solution with 100$\%$ of the shots and noise-free input—we observe a smaller entry count ratio for higher frequency indices (e.g., those greater than 120) in Figure~\ref{fig:entryratio50}. This result is expected, as the benchmark solution derived from noisy and reduced input is inherently noisier compared to the solution obtained with all noise-free shots. The induced noise in the benchmark solution lacks low-rank properties, which contributes to the observed difference.

The results produced by the proposed MDD solver under the noisy and reduced-shot conditions are shown in Figure~\ref{fig:t_x_tlr_perc50}. While the benchmark solver introduces noticeable noise and artifacts under these conditions, the proposed solver produces results that remain consistent with those in Figure~\ref{fig:t_x_tile64_20diags_r3}, which were obtained using all shots without noise. Similarly, Figure~\ref{fig:entryratio25} presents the entry count ratio analyzed from the benchmark solution with noisy input and only 25$\%$ of the total shots. As seen in the zoomed-in plot, the entry count ratio is smaller compared to that in Figure~\ref{fig:entryratio50}, indicating that the noise in the benchmark solution, due to further reduced shots and added noise, increases the local ranks.

Figures~\ref{fig:t_x_tik_perc25} and \ref{fig:t_x_tlr_perc25} illustrate the results from the benchmark solver and the proposed solver, respectively, when only 25$\%$ of the 26,040 shots are used, with Gaussian noise added to the input data. In this case, the noise in the solution produced by the benchmark solver is clearly visible. These results demonstrate that the proposed solver is more robust to sparse shot sampling and less sensitive to noise compared to the benchmark solver.

In summary, we have demonstrated that the proposed MDD solver exhibits strong robustness to sparse shot sampling and random noise, outperforming the benchmark solver in terms of de-multiple performance and signal-to-noise ratio. One might question whether the benchmark solver’s performance could be improved by simply increasing the regularization parameter $\lambda$ in Equation~\ref{eq:bench_solver} (the Tikhonov-type regularization term). To address this, we tested three values for $\lambda$ ($1.0$, $0.1$, and $0.01$) and found that increasing $\lambda$ in the benchmark solver did not effectively mitigate the noise.

\section{Discussion}
The proposed adaptive tile low-rank factorization relies on the solution from the benchmark solver to determine the optimal tile size and local ranks. However, the benchmark solution is sensitive to noise and shot sampling. Therefore, exploring strategies to determine the optimal tile size and local ranks without using the benchmark solver could further enhance the robustness of the proposed method.

Additionally, our method could benefit from leveraging the tile low-rank matrix-matrix multiplication algorithm to reduce computational cost. This is an area of ongoing research.

\section{Conclusions}
In this paper, we propose a novel MDD solver to address the challenges of reconstructing Green's functions in large-scale 3D seismic data inversion. The solver leverages adaptive tile-based low-rank factorization and Hilbert space-filling curve reordering to efficiently reduce the dimensionality of the unknown Green's function matrix, improving its robustness and accuracy. Through extensive numerical tests using the 3D EAGE/SEG Overthrust model, we demonstrate that our method offers significant improvements in both signal-to-noise ratio and de-multiple performance compared to traditional inversion methods.

The results show that, by adaptively adjusting tile sizes based on frequency, the solver effectively handles the varying low-rank properties across different frequencies, providing a more accurate reconstruction of the Green's function. Our method also eliminates the artifacts observed in solutions obtained from global low-rank factorization, which can suffer from residual free-surface multiples and noise. Moreover, the proposed solver maintains its robustness even under challenging conditions such as sparse shot sampling and noisy input data. In these cases, it outperforms benchmark solvers by producing cleaner and more stable results.

This work highlights the potential of adaptive low-rank factorization techniques in large-scale 3D seismic inversion problems The proposed MDD solver offers a promising approach to tackle complex seismic data inversion tasks, paving the way for its broader application in geophysical exploration and related fields.

\section{ACKNOWLEDGMENTS}
We thank King Abdullah University of Science \& Technology (KAUST) for supporting this research. For computer time, this research leveraged the resources of the Supercomputing Laboratory at KAUST in Thuwal, Saudi Arabia. We thank Yuxi Hong and Daria Sushnikova for their useful discussions. We also acknowledge the use of ChatGPT in the process of writing the paper.
\newpage
\bibliographystyle{seg}  
\bibliography{example}

\end{document}